\documentclass[lettersize,journal]{IEEEtran}
\usepackage{amsmath,amsfonts}
\usepackage{amssymb}
\usepackage{algorithmic}
\usepackage{algorithm}
\usepackage{array}
\usepackage[caption=false,font=normalsize,labelfont=sf,textfont=sf]{subfig}
\usepackage{textcomp}
\usepackage{stfloats}
\usepackage{url}
\usepackage{verbatim}
\usepackage{graphicx}
\usepackage{cite}
\usepackage{booktabs}
\usepackage{multirow}
\usepackage{makecell} 
\usepackage{fontawesome}
\usepackage{pifont}
\begin{document}

\title{Physics-informed Diffusion Models for Multi-scale Prediction of Reference Signal Received Power in Wireless Networks}

\author{
    Xiaoqian Qi, Haoye Chai\IEEEauthorrefmark{1},~\IEEEmembership{Member,~IEEE}, Yue Wang, Zhaocheng Wang,~\IEEEmembership{Fellow,~IEEE}, \\and Yong Li,~\IEEEmembership{Member,~IEEE}
    
    \thanks{X. Qi, Y. Wang, Z. Wang, and Y. Li are with the Department of Electronic Engineering, BNRist, Tsinghua University, Beijing 100084, China.}
    \thanks{H. Chai is with the State Key Laboratory of Networking and Switching Technology, Beijing University of Posts and Telecommunications, Beijing, China (E-mail: haoyechai@bupt.edu.cn).}
    \thanks{\IEEEauthorrefmark{1} Corresponding author.}
    
}

\markboth{IEEE TRANSACTIONS ON MOBILE COMPUTING}%
{Shell \MakeLowercase{\textit{et al.}}: A Sample Article Using IEEEtran.cls for IEEE Journals}


\maketitle

\begin{abstract}
The Reference Signal Received Power (RSRP) is a crucial factor that determines communication performance in mobile networks. Accurately predicting the RSRP can help network operators perceive user experiences and maximize throughput by optimizing wireless resources. However, existing research into RSRP prediction has limitations in accuracy and verisimilitude. Theoretical derivations and existing data-driven methods consider only easily quantifiable Large-Scale (LS) information, and struggle to effectively capture the intertwined LS and Small-Scale (SS) signal attenuation characteristics of the wireless channel. Moreover, the lack of prior physical knowledge leads to weak accuracy, interpretability, and transferability. In this paper, we propose a novel RSRP prediction framework, Channel-Diff. This framework physically models LS and SS attenuation using multimodal conditions and employs physics-informed conditional diffusion models as the prediction network. Channel-Diff extracts prior physical information that characterises the signal propagation process from network parameters and multi-attribute maps of the urban spatial environment. It provides LS physical priors through large-scale propagation modelling and shadow–occlusion modelling, and SS physical priors through multipath propagation modelling and urban microenvironment feature extraction. We design a physical-prior-guided two-stage training scheme with a noise prior guidance mechanism, enabling effective fusion of multi-scale physical knowledge with the diffusion models. Evaluations demonstrate Channel-Diff exhibits excellent performance on RSRP prediction, achieving at least 25.15\%–37.19\% improvement in accuracy relative to baseline methods. Additionally, the model also demonstrated outstanding performance in terms of transferability and training efficiency.
\end{abstract}

\begin{IEEEkeywords}
Reference signal received power prediction, Diffusion models, Physical-informed generation, Channel estimation.
\end{IEEEkeywords}

\section{Introduction}
\IEEEPARstart{R}{eference} signals are deterministic signals transmitted by the Base Station (BS) to User Equipment (UE) for specific reference purposes, such as synchronization and Channel State Information (CSI) estimation. The Reference Signal Received Power (RSRP) reflects the quality of the wireless channel and is used by network operators to estimate coverage and evaluate user communication experience. In current mobile networks, RSRP is typically obtained through measurement report feedback from the UE in real time, after which the BS adjusts the network status in the next time slot. Given the increasing demands for large-scale, low-latency communication, the approach introduces processing delays and struggles to keep up with rapidly changing network dynamics.

An effective solution is to predict the RSRP in advance. In November 2024, the 3rd Generation Partnership Project (3GPP) RAN4 \#106 meeting~\cite{du2024ai} explicitly highlighted the need for RSRP prediction algorithms. It is noted that RSRP prediction is a key component of indirect measurement event prediction and A3 event (i.e., handover decision event) prediction, and can be used to optimize network handovers~\cite{ho-10477096, ho-10654787, ho-9762537}. In addition, RSRP prediction can also provide an important reference for tasks such as cell selection~\cite{cs-10.1145/3544017, cs-9789832, cs-GURES2023621}, resource allocation~\cite{ra-9956752, ra-cpe6228, ra-electronics12204209}, and power control~\cite{pc-2024arXiv240513653J, pc-9326357}. With accurate RSRP prediction techniques, operators can proactively assess network coverage and design network optimization plans even before deployment. RSRP prediction can also empower the simulation of wireless channels in wireless network digital twins~\cite{10.1145/3711682, 2025arXiv250312177I,VAE-9978065,2024arXiv241013379Y, 2024arXiv240108023Z, 2024arXiv241022437S, 2025arXiv250108680L}, assisting in the construction of a world model where communication systems serve as the information backbone. This will enable governments and operators to simulate communication networks and perform counterfactual reasoning without relying on already-deployed infrastructure, providing a forward-looking perspective for policy-making.

Existing RSRP prediction methods can be categorized into three approaches: formula-based calculation, system-level simulation, and Machine Learning (ML) models. Formula-based methods estimate the Path Loss (PL) of signal power using propagation models. Classical propagation models include Okumura~\cite{y__okumura_1968}, Hata~\cite{Hata} and WINNER II~\cite{WINNERII}, which differ mainly in the applicable frequency bands and scenarios (e.g., urban, rural). Due to the fixed parameters, these models are unable to account for environmental variability within the same type of scenario. With the advancement of simulation tools, RSRP can be estimated through system-level simulations, such as NS-3~\cite{NS3-R} and Omnet~\cite{10.5555/1416222.1416290}. Compared with propagation models, simulation models allow for more detailed definitions of scenarios and material properties. However, they introduce a large computational overhead and rely on idealized channel conditions, which limit their realism.

ML offers more flexible schemes for RSRP prediction by enabling models to learn complex shared patterns from large volumes of data. Wu~\textit{et al.}~\cite{GBDT-9700950} employed Gradient Boosting Decision Trees (GBDT) to perform regression on RSRP using a combination of features such as BS-UE distance and antenna tilt. Yu~\textit{et al.}~\cite{10597656} proposed a federated learning algorithm based on Open-Radio Access Network (O-RAN) and designed a multi-head Deen Neural Network (DNN) to mitigate the performance loss caused by user and environmental diversity. Li~\textit{et al.}~\cite{VAE-9978065} developed a data-driven two-layer neural network that first uses a Variational Autoencoder (VAE) to extract environment-related features associated with UE, BS, and network Key Performance Indicators (KPIs), and then applies a likelihood model to predict RSRP. However, existing ML models exhibit two main limitations:

\subsubsection{\textbf{Insufficient consideration of the multi-scale channel characteristics}} In wireless channels, signal power attenuation can be divided into two components: Large-Scale (LS) attenuation and Small-Scale (SS) attenuation~\cite{GRAMI2016493,KALUUBA2006621,QUEIROZ201796}. LS attenuation arises from PL due to radiative dispersion and shadowing effects caused by obstructions such as buildings, while SS attenuation is caused by factors like multipath propagation, Doppler shifts due to mobile terminal movement, as well as rapid environmental fluctuations. These factors are highly complex and entangled. What's more, SS attenuation is typically influenced by environmental features that are difficult to quantify. Most ML-based models consider only LS factors, which are determined by BS parameters and the relative position between the BS and the UE. They seldom consider the environmental features associated with SS attenuation. This leads to a loss of scale when measuring the channel state.

\subsubsection{\textbf{Lack of guidance from prior physical knowledge}} Propagation models are grounded in electromagnetic propagation theory and empirical analysis, revealing explicit mathematical relationships between PL and multiple LS factors. However, existing ML models are data-driven and struggle to identify such physical laws due to the entangled and complex nature of environmental and multi-scale features. This limitation leads to poor interpretability and weak generalization across different scenarios. Further, this makes models prone to falling into suboptimal solutions that deviate from physical reality.

To address the above limitations, we propose a novel RSRP prediction framework, Channel-Diff. This framework physically models LS and SS attenuation using multimodal conditions and employs physics-informed conditional diffusion models as the prediction network. \textbf{First}, it adopts a conditional diffusion model as the backbone, considering multi-modal conditions from two scales. For LS conditions, we select 10 network parameters determined by the BS, the UE, and the relative position between them. For SS conditions, we construct a spatial multi-attribute urban environment map that contains the building height and ground altitude distribution. We customize the denoising network to effectively control the RSRP prediction sequence with the selected conditions. \textbf{Second}, we propose physics-informed conditional diffusion models that integrate prior physical knowledge and physical representations of LS and SS attenuation into the neural networks. Unlike PINN~\cite{RAISSI2019686}, which represents physical processes using differential equations, we are inspired by knowledge distillation~\cite{2015arXiv150302531H, Karniadakis2021} and introduce a physics-prior-guided two-stage training paradigm to infuse the diffusion model with LS physical knowledge. We further propose a noise-prior guidance mechanism based on occlusion and shadow modelling to prevent knowledge forgetting during the two-stage training process. \textbf{Third}, we provide multi-scale physical representations and feature learning methods for the RSRP prediction problem. Based on the multi-attribute map, we compute an occlusion factor via occlusion and shadow modelling for noise-prior guidance, and construct a reflection embedding via multipath propagation modelling for learning SS multipath effects. We further design a Microenvironment Feature Extraction Network (MFEN), leveraging edge enhancement and spatial attention mechanisms to learn SS attenuation induced by the surrounding environment. 

We summarize our contributions from three aspects:
\begin{itemize}
    \item We propose physics-informed conditional diffusion models for RSRP prediction. The physics-prior-guided two-stage training paradigm and prior-noise guidance effectively inherit and connect physical knowledge across stages, enabling the model to jointly learn LS and SS attenuation characteristics, achieving high accuracy and interpretability.

    \item To the best of our knowledge, this is the first generation-based RSRP prediction model that utilizes physical models of wireless propagation to construct representative embeddings. We build LS feature representations through large-scale propagation and occlusion\&shadow modelling, and SS feature representations through multipath propagation modelling and the MFEN. This provides multi-dimensional physical priors for RSRP prediction.

    \item We evaluate the effectiveness of Channel-Diff on two real-measured RSRP datasets. Channel-Diff achieves performance improvements of 37.19\% and 25.15\% over the second-best models on the two datasets, respectively. Moreover, we demonstrate that the incorporation of physical knowledge significantly enhances model performance and validate the advantages of Channel-Diff in terms of transferability and training efficiency.
\end{itemize}


\section{Related Works}
\label{related_works}
\subsubsection{Machine Learning Based RSRP Prediction}
ML has demonstrated advantages over traditional statistical methods in identifying complex associations between RSRP and multiple influencing factors, as well as in capturing intricate environmental features embedded in data. Tao~\textit{et al.}~\cite{BPNN} selected features based on the Cost 231-Hata model and employed a Backpropagation Neural Network (BPNN) to construct an intelligent wireless propagation model. Kanto~\textit{et al.}~\cite{LSTM-10575638} utilized a Long Short-Term Memory (LSTM) network to leverage the sequential information and capture the temporal characteristics of RSRP. Wu~\textit{et al.}~\cite{GBDT-9700950, 9674686} combined a Gradient Boosting Decision Tree (GBDT)-based feature generator with DNNs to predict RSRP, demonstrating the complementary benefits of GBDT for neural network-based learning. Li~\textit{et al.}~\cite{VAE-9978065} designed a data-driven two-stage neural model that first extracts environmental features via a VAE. \cite{EM-10745208, EM-2024arXiv240808593W, EM-9354041, EM-9523765, EM-9931518, EM-10227351} reformulate RSRP prediction as an image generation problem, leveraging ML models’ strong image processing capabilities to construct RMs. Zheng~\textit{et al.}~\cite{EM-10227351} proposed RadioGAN, a Generative Adversarial Networks (GANs)-based model for radio map construction, which uses building distribution maps as priors in the generator and incorporates a propagation model and spatial masks into the discriminator, achieving strong performance on real-world datasets. Wang~\textit{et al.}~\cite{EM-2024arXiv240808593W} proposed RadioDiff, a diffusion-model-based RM constructor. It employs a U-Net architecture and guides the denoising process using vehicle distributions, building distributions, and base station locations.

\subsubsection{Multivariate Series Generation}
GANs have shown significant performance in multivariate series generation~\cite{10461110, 9685707, 10.1016/j.phycom.2023.102214}. Madane~\textit{et al.}~\cite{MTS-CGAN-2022arXiv221002089M} proposed MTS-CGAN, a conditional GAN capable of modelling real multivariate time series under various types of conditions. Yoon~\textit{et al.}~\cite{TimeGAN-10.5555/3454287.3454781} introduced TimeGAN, which combines the flexibility of an unsupervised framework with the control of supervised training, achieving remarkable performance. Compared to GANs, diffusion models generate data through a step-by-step denoising process, offering better training stability and enhanced capability for multi-pattern data generation~\cite{10.5555/3540261.3540933}. Tashiro~\textit{et al.}~\cite{CSDI} proposed CSDI, a score-based diffusion model designed for multivariate time series imputation. Peebles~\textit{et al.}~\cite{10377858} introduced DiT, a Transformer-based diffusion model that incorporates conditioning information via layer normalization and transforms the noise loss into a variable loss, enabling more flexible loss design. Narasimhan~\textit{et al.}~\cite{10.5555/3692070.3693584} designed Time Weaver, an innovative diffusion architecture that significantly improves time series generation by incorporating heterogeneous metadata in the form of categorical, continuous, and even time-varying variables. However, few studies have explored the use of GANs or diffusion models for RSRP prediction.

\subsubsection{Physics-informed Machine Learning}
Physics-Informed Machine Learning (PIML) has gained significant attention for its ability to incorporate physical laws into machine learning models~\cite{Karniadakis2021}. By combining the power of data-driven approaches with the constraints of physics, PIML offers a promising solution to problems where labelled data is scarce or expensive to obtain. Raissi \textit{et al.}~\cite{RAISSI2019686} first introduced Physics-Informed Neural Networks (PINNs) in 2019, which embed the governing equations of physical systems into the training process of neural networks. As a combination of PIML and diffusion models, Bastek \textit{et al.}~\cite{2024arXiv240314404B} informs the diffusion models of potential constraints during model training. Knowledge distillation~\cite{2015arXiv150302531H, Gou2021} is a technique used to transfer knowledge from a larger, complex model (the teacher) to a smaller, simpler model (the student). Tee \textit{et al.}~\cite{2024arXiv241108378T} introduced Physics-Informed Distillation (PID) into diffusion models, conceptualizing diffusion models as ODE systems. Zhang~\textit{et al.}~\cite{10.1145/3534678.3539440} incorporated a gravity model into a diffusion framework, designing a novel neural network model alongside a physics-informed crowd simulator.

\section{Preliminaries and Problem Formulation}
\label{sec_Pre&Prob}

\subsection{Multi-scale Characteristics of Wireless Signal Propagation}
\label{subsec:factors}
\begin{figure*}[tb]
\centering
\includegraphics[width=0.85\linewidth]{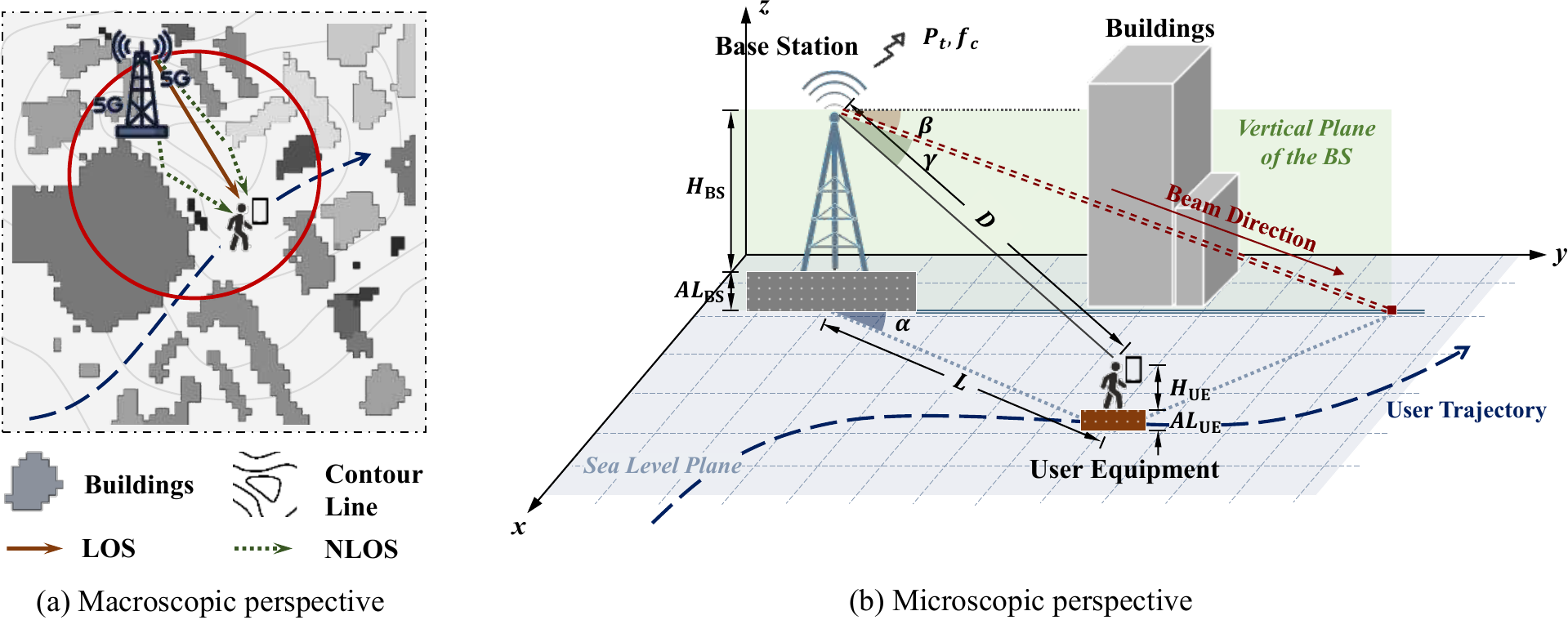}
\caption{A typical scenario of wireless transmission. (a) provides a macroscopic perspective, which shows the user's surrounding urban environment. The shaded areas represent buildings, where the shading intensity indicates building height. The contour lines denote variations in ground altitude. The LOS path and several NLOS paths are illustrated in the figure. (b) provides a microscopic perspective, which defines the relevant network parameters and beam ray information.}
\vspace{-4mm}
\label{Fig_trans}
\end{figure*}

Wireless signal propagation in urban spaces experiences power attenuation from the Reference Signal Transmission Power (RSTP or RSP) to the RSRP due to multiple factors. This power attenuation is mainly composed of two parts: the Large-Scale (LS) attenuation caused by PL, occlusion and shadowing, as well as the Small-Scale (SS) attenuation caused by multipath propagation and BS-UE relative motion. Fig.~\ref{Fig_trans} shows the scenario of a mobile user moving through the urban space. Signal transmission between the BS and UE consists of Line-Of-Sight (LOS), i.e., the direct path, and Non-Line-Of-Sight (NLOS), which is mainly composed of reflected paths. Table~\ref{Tab_factors} lists the key factors related to RSRP determination in LOS scenarios and provides brief explanations.

\begin{table}[!h]
\centering
\caption{Key factors related to RSRP determination in LOS scenarios.}
\begin{tabular}{c c p{4cm}} 
\toprule
\textbf{Parameter Type} & \textbf{Symbol} & \textbf{Description} \\
\midrule
\multirow{7}{*}{BS parameters}
    & $H_{\text{BS}}$ & height of BS antenna from ground \\
    & $AL_{\text{BS}}$ & altitude of the BS tower base \\
    & $h_t$ & $H_{\text{BS}}+AL_{\text{BS}}$, altitude of the BS antenna\\
    & $P_t$ & transmit power \\
    & $f_c$ & carrier frequency \\
    & $\gamma$ & beam downtilt angle \\
\midrule
\multirow{5}{*}{UE parameters}
    & $H_{\text{UE}}$ & height of UE antenna from the ground \\
    & $AL_{\text{UE}}$ & ground altitude of the user \\
    & $h_r$ & $H_{\text{UE}}+AL_{\text{UE}}$, altitude of the UE antenna\\
\midrule
\multirow{8}{*}{Relative position}
    & $L$ & 2D BS-UE distance \\
    & $D$ & 3D BS-UE distance \\
    & $N_b$ & the number of buildings between the BS and UE along the LOS \\
    & $\alpha$ & aerial angle between the beam direction and the LOS \\
    & $\gamma$ & ground angle between the beam direction and the LOS \\
\bottomrule
\end{tabular}
\label{Tab_factors}
\end{table}

The received signal is typically expressed as $\pmb{x}_r = H \pmb{x}_t + n$, where $\pmb{x}_t$ denotes the transmitted signal. $H$ is the time-domain channel impulse response matrix, which can be expressed as $H = h_{0} + \sum_{i=1}^{N_{\mathrm{NLOS}}} h_i$. Here, $h_0$ and $h_i~(i>0)$ represent the channel responses of the LOS path and $N_{\mathrm{NLOS}}$ reflected paths, respectively. $n$ is random noise. Therefore, 
\begin{equation}
\small
    \pmb{x}_r = \left( h_{0} + \sum_{i=1}^{N_{\mathrm{NLOS}}} h_i \right) \pmb x_t + n = h_{0} \pmb x_t  + \sum_{i=1}^{N_{\mathrm{NLOS}}} h_i \pmb x_t + n.
\end{equation}

LS attenuation is primarily determined by the LOS path between the BS and the UE, i.e., it affects the received signal through $h_{0}$. Considering only antenna configuration and PL, the received power $P_r$ can be formulated as
\begin{equation}
\small
    P_r=P_t+G_t+G_r-\text{PL} + u,
\label{eq_PR}
\end{equation}
where $G_t$ is the transmit antenna gain, $G_r$ is the receive antenna gain and $u$ is a residual term. When LOS is absent, blockage and shadowing become the dominant factors. The $1^\text{st}$ Fresnel zone is defined as the region between the BS and UE in which the path-length difference does not exceed half a wavelength; obstacles within this region can cause significant diffraction and attenuation effects on the signal.


SS attenuation is caused by frequency-selective fading induced by multipath effects and time-selective fading induced by the Doppler effect. The Doppler effect can be mathematically represented as a multiplicative perturbation on $h_i~(i \geq 0)$. It is related to factors such as UE mobility and the velocities of surrounding vehicles, which are highly random and typically difficult to model. Multipath effects influence the received signal by affecting $h_i~(i>0)$. Assuming that path $i$ undergoes $n$ reflections in total, the complex coefficient $h_i$ can be conceptually written as
\begin{equation}
\small
    h_i \propto \frac{1}{d_{\mathrm{ref}}^{\eta}} \cdot \prod_{j=1}^{n} \rho_j \cdot e^{-j2\pi d_{\mathrm{ref}}/\lambda},
\end{equation}
where $d_{\mathrm{ref}}$ is the total propagation distance, $\eta$ is a PL exponent, $\rho_j$ denotes the reflection-related attenuation at the $j$-th bounce, and $\lambda=c/f_c$ is the wavelength. As a result, even when the LS term is similar, the coherent superposition in $H$ may be constructive or destructive, leading to rapid SS fluctuations in received power. For the SS attenuation modelling in this paper, we consider only the multipath effects, which originate from multi-order reflection paths caused by reflective surfaces such as the ground and buildings.

\subsection{Conditional Diffusion Models}
\label{sec::CDM}
Diffusion models were originally proposed by Sohl-Dickstein \textit{et al.}~\cite{DBLP} in 2015 and have since evolved into widely known generative models such as DDPM~\cite{DDPM} and CSDI~\cite{CSDI}. The core idea of diffusion models involves hypothetically defining a forward process that gradually adds noise to the data, and then recovering the target data by estimating and removing the noise in the reverse process. The forward process contains a series of adding-noise steps, gradually transforming data samples towards a Gaussian distribution with zero mean and unit covariance. Given a data sample $\pmb{x}_0^{\text{ta}}$ from the target space $\pmb{\mathcal{X}}^{\text{ta}}$, the forward process is represented as a Markov chain with each step defined by:
\begin{equation}
\small
\begin{aligned}
    q(\pmb{x}^{\text{ta}}_{1:K}|\pmb{x}^{\text{ta}}_0) &:= \prod_{k=1}^K q(\pmb{x}^{\text{ta}}_k|\pmb{x}^{\text{ta}}_{k-1}), \\
    q(\pmb{x}^{\text{ta}}_k|\pmb{x}^{\text{ta}}_{k-1}) &:= \mathcal{N}(\sqrt{1-\beta_k}\pmb{x}^{\text{ta}}_{k-1}, \beta_k \pmb{I}),
\end{aligned}
\label{eq:forward_process}
\end{equation}
where $\beta_k$ controls the noise intensity at each step. Defining $\hat{\alpha}_k := 1-\beta_k$ and $\alpha_k := \prod_{i=1}^k \hat{\alpha}_k$, the data sample in step $k$ can be reformulated as $\pmb{x}^{\text{ta}}_k = \sqrt{\alpha_k}\pmb{x}^{\text{ta}}_0 + (1-\alpha_k)\pmb\epsilon$, where $\epsilon \sim \mathcal{N}(\pmb{0}, \pmb{I})$. It indicates that the data at any step is a linear combination of the initial data sample $\pmb{x}^{\text{ta}}_0$ and noise $\pmb\epsilon$.

The reverse process aims to iteratively restore data samples from noise at each step. Consider a condition $\pmb{x}^{\text{co}} \in \pmb{\mathcal{X}}^{\text{co}}$, the conditional reverse process can be formulated as a Markov chain with each step defined by:
\begin{equation}
\small
\begin{aligned}
    &p_\theta(\pmb{x}^{\text{ta}}_{0:K} |\pmb{x}^{\text{co}}) := p(\pmb{x}^{\text{ta}}_K) \prod_{k=1}^K p_\theta(\pmb{x}^{\text{ta}}_{k-1}|\pmb{x}^{\text{ta}}_k, \pmb{x}^{\text{co}}), \\
    &p_\theta(\pmb{x}^{\text{ta}}_{k-1}|\pmb{x}^{\text{ta}}_k, \pmb{x}^{\text{co}}) := \mathcal{N}(\pmb{x}^{\text{ta}}_{k-1}; \mu_\theta(\pmb{x}^{\text{ta}}_k, k|\pmb{x}^{\text{co}}), \sigma_\theta(\pmb{x}^{\text{ta}}_k, k|\pmb{x}^{\text{co}})\pmb{I}),
\end{aligned}
\label{eq:reverse_process}
\end{equation}
where $\mu_\theta$ and $\sigma_\theta$ are trainable parameters representing the mean and variance of the model distribution at each step. Ho \textit{et al.}~\cite{DDPM} formulated the optimization objective of the denoising network $\pmb{\epsilon}_\theta$ as follows:
\begin{equation}
\small
    \min_\theta \mathcal{L}(\theta) := \min_\theta \mathbb{E}_{\pmb{x}^{\text{ta}}_0 \sim q(\pmb{x}^{\text{ta}}_0), \pmb{\epsilon} \sim \mathcal{N}(\pmb{0},\pmb{I}), k} \left\| \pmb\epsilon - \pmb{\epsilon}_\theta(\pmb{x}^{\text{ta}}_k, k|\pmb{x}^{\text{co}}) \right\|_2^2,
\label{eq:objective}
\end{equation}
where $\pmb{x}^{\text{ta}}_k = \sqrt{\alpha_k}\pmb{x}^{\text{ta}}_0 + (1-\alpha_k)\pmb{\epsilon}$. During each step $k$, the model predicts the denoised output of $\pmb{x}_k^{\text{ta}}$ by estimating $\mu_\theta$ and $\sigma_\theta$, thereby progressively recovering samples that match the target distribution under the given condition.

\subsection{Problem Formulation}
\label{subsec::problem_def}

\noindent\textbf{Problem Definition.} \textit{Given the network parameters and urban environment context, predict the UE-side RSRP series aligned with the user's mobile trajectory.}

The RSRP prediction problem defined in this paper is a conditional generation task over multi-feature series. Unlike research such as static electromagnetic map reconstruction, we adopt a user-centric perspective, aiming to estimate channel quality information along a continuous trajectory using generative methods. We consider a user trajectory $\pmb{\Gamma}^{1 \times T}$ consisting of $T$ consecutive time steps, and our goal is to estimate the RSRP value at each trajectory point by leveraging multi-scale channel conditions and environmental context aligned with $\pmb{\Gamma}$, yielding $\pmb{x} = (x^{(1)}, x^{(2)}, \ldots, x^{(T)})^\top$.

To ensure the generalizability of RSRP prediction, the model should possess the following fundamental capabilities. First, it should exhibit the generic competence of a time-series generative model, that is, the ability to predict a target sequence conditioned on multimodal inputs while capturing the intrinsic temporal dependencies within the series. Second, it should incorporate physical knowledge of signal propagation in urban electromagnetic environments, including large-scale path loss and building-induced blockage, as well as small-scale fading characteristics arising from multipath effects. Therefore, we formulate the problem as:
\begin{equation}
\small
    \pmb{x} = \mathcal{P} \left(\pmb{x}^{\text{co}}\right),
\end{equation}
where $\mathcal{P}$ denotes the RSRP prediction model, and $\pmb{x}^{\text{co}}$ represents the external multi-scale conditions characterizing the channel states.

\section{System Model}
\label{method}
\begin{figure*}[tb]
\centering
\includegraphics[width=0.9\linewidth]{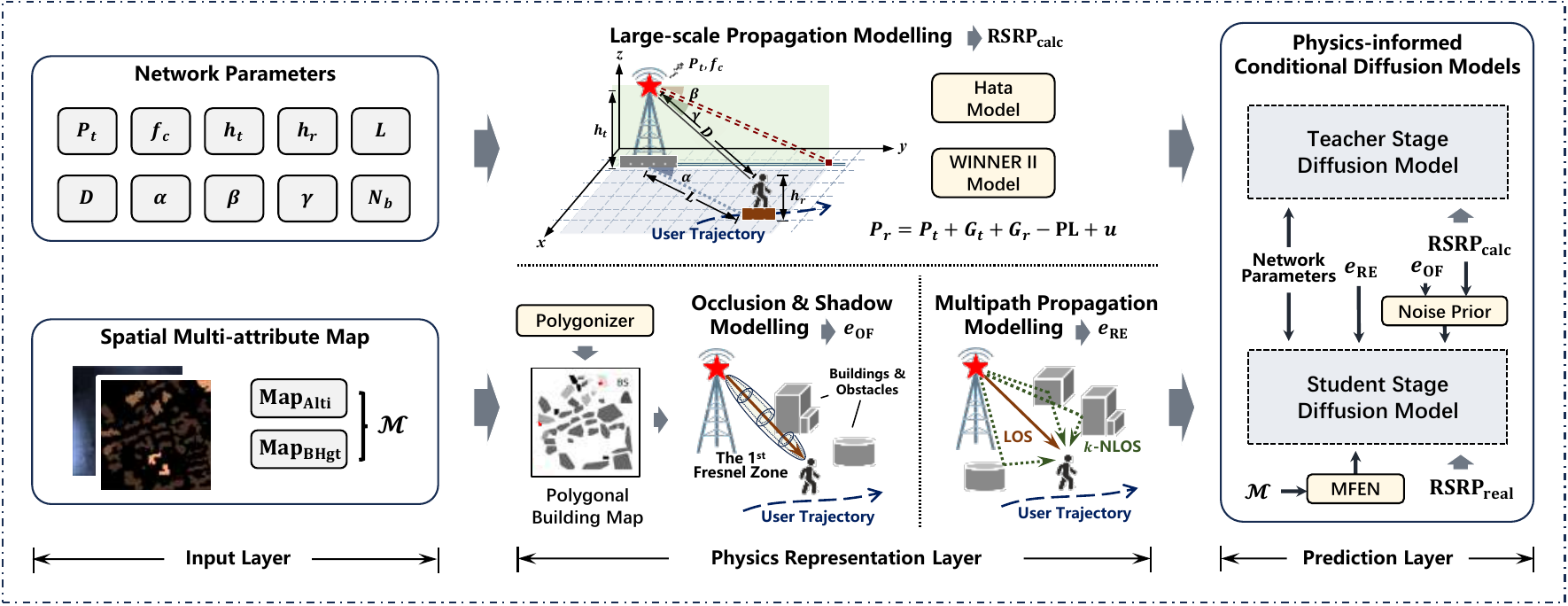}
\caption{Overall framework of Channel-Diff.}
\label{Fig_syst}
\vspace{-4mm}
\end{figure*}

To solve the problem, we propose a novel framework, Channel-Diff, which infuses the physical knowledge of multi-scale signal propagation into custom-designed physics-informed diffusion models. This three-layer framework is shown in Fig~\ref{Fig_syst}.

\subsubsection{{Input Layer}} The input layer contains two main types of multi-modal external conditions, including network parameters and a spatial multi-attribute map. The network parameters form a 10-dimensional sequence-modality conditional set, including $P_t,~f_c,~h_t,~h_r,~L,~D,~\alpha,~\beta,~\gamma,~\text{and}~N_b$. The spatial multi-attribute map is a graph-modality representation of the urban environmental context, consisting of the ground elevation map $\text{Map}_\text{Alti}$ and the building-height distribution map $\text{Map}_\text{BHgt}$. We define the urban multi-attribute map as $\mathcal{M}=[\text{Map}_\text{Alti}, \text{Map}_\text{BHgt}]$. It represents the urban microenvironment required for modelling blockage, shadowing, and multipath propagation, as well as for extracting SS micro-environment features around the UE.

\subsubsection{{Physics Representation Layer}} This layer consisted of three physical modelling modules. The Large-scale Propagation Modelling (LPM) module is based on propagation models including the Hata model and the WINNER II model, and outputs a calculated target value, $\text{RSRP}_\text{calc}$, which mainly considers the theoretical LOS power attenuation caused by LS signal propagation. Due to image resolution limitations, buildings in $\mathcal{M}$ exhibit jagged edges, which can introduce significant errors in signal reflection modelling. Without loss of generality, we apply a polygonizer to simplify contiguous building regions into polygons with 3~6 edges, yielding a polygonal building map that is used for Occlusion \& Shadow Modelling (OSM) and Multipath Propagation Modelling (MPM). These three modules enable multi-scale physical modelling of the wireless channel. LPM and OSM focus on LS attenuation characteristics, capturing PL caused by propagation distance as well as blockage and shadowing induced by building distributions. MPM models SS attenuation characteristics associated with multipath propagation. These modules provide the prior physical knowledge for the multi-scale prediction of RSRP.

\subsubsection{{Prediction Layer}} We propose Physics-informed Conditional Diffusion Models to serve as the prediction layer of Channel-Diff. This module adopts a two-stage training scheme to incorporate the physical knowledge modelled in the physics representation layer into the training process of the diffusion models. We will provide a detailed description of this layer in Section~\ref{sec:PIDM}.

\subsection{Large-scale Propagation Modelling}
\label{sec::LS_Modelling}
Propagation models are based on electromagnetic theory to model the power loss during long-distance signal transmission. The first propagation model is the Free Space Propagation Model (FSPM)~\cite{1697062}. It assumes an ideal environment with no obstacles, reflections, diffractions, or scattering effects along the propagation path. After that, researchers developed numerous measurement-based propagation models in different environments and frequency bands. Hata~\cite{Hata} and WINNER II~\cite{WINNERII} are two of the most widely used propagation models, applicable to different frequency bands. Hata model is applicable in $150\sim1500$ MHz, while WINNER II~\cite{WINNERII} is applicable in $2\sim6$ GHz. Measurement-based propagation models are functions of PL and network parameters, and can be expressed as
\begin{equation}
\small
    \text{PL}=f(D,~f_c,~h_t,~h_r).
\end{equation}
When calculating RSRP, different propagation models need to be selected according to the BS’s frequency band. As introduced in Section~\ref{subsec:factors}, the received power $P_r$ is determined by Eq.~(\ref{eq_PR}). Finally, aligned with the user trajectory $\pmb{\Gamma}$, $\textbf{RSRP}_\text{calc}$ can be represented by
\begin{equation}
\small
    {\textbf{RSRP}_\text{calc}}^{1 \times T}=[P_r^{(0)}, P_r^{(1)}, ..., P_r^{(T)}]^\top.
\end{equation}

\subsection{Occlusion \& Shadow Modelling}
Propagation models such as Hata and WINNER II are primarily designed for LOS scenarios, thus often fail to accurately capture the actual signal attenuation in urban NLOS or heavily blocked environments. To characterize the reliability of $\mathbf{RSRP}_\mathrm{calc}$ under real propagation conditions, we construct a normalized occlusion factor $\pmb{e}_{\mathrm{OF}}$. As shown in Fig.~\ref{Fig_syst}, the signal may be intruded upon by building rooftops or edges entering the $1^\text{st}$ Fresnel Zone, thereby introducing additional diffraction loss. For any candidate blocking path, we first determine the intersection point $O_\text{LOS}$ between the BS–UE line and the building footprint in the horizontal plane, and compute the corresponding horizontal distances from this point to the BS and UE, denoted $d_{O_\text{LOS}-\text{BS}}$ and $d_{O_\text{LOS}-\text{UE}}$. The radius of the $1^\text{st}$ Fresnel zone is given by
\begin{equation}
\small
    r_{\mathrm{F,1}} = \sqrt{\frac{\lambda d_{O_\text{LOS}-\text{BS}} d_{O_\text{LOS}-\text{UE}}}{d_{O_\text{LOS}-\text{BS}} + d_{O_\text{LOS}-\text{UE}}}}.
\end{equation}
Let $h_{O_\text{LOS}}$ be the unobstructed ray height along the BS–UE line at the intersection, and $h_{\mathrm{BHgt}}$ the building rooftop height. Then the intrusion height into the first Fresnel zone is $\Delta h = h_{\mathrm{BHgt}} - h_{O_\text{LOS}}$. When $\Delta h > 0$, the building penetrates the $1^\text{st}$ Fresnel zone and causes diffraction loss; when $\Delta h \leq 0$, its diffraction impact on this path is negligible. Based on the classical knife-edge diffraction model, we define the diffraction parameter
\begin{equation}
\small
    v = \sqrt{2}\,\frac{\Delta h}{r_{\mathrm{F,1}}} ,
\end{equation}
and use an empirical formula as Eq.~(\ref{eq:L_d_v}) to obtain the additional diffraction loss at the blocking point:
\begin{equation}
\small
    L_d(v) =
    \begin{cases}
        0, & v \leq -0.7,\\[2pt]
        6.9 + 20\log_{10}\!\Big(
            \sqrt{(v-0.1)^2 + 1} + v - 0.1
        \Big), & v > -0.7,
    \end{cases}
\label{eq:L_d_v}
\end{equation}
where $L_d(v$) is expressed in dB.

Suppose that multiple buildings lie between the BS and UE. For the $j$-th building intersection point $O_{\text{LOS},j}$, the radius of the $1^\text{st}$ Fresnel zone is denoted by $r_{\mathrm{F},1;j}$. So, the knife-edge dimensionless parameter is given by $v_j=\sqrt{2},\Delta h / r_{\mathrm{F},1;j}$, and the diffraction loss is $L_d(v_j)$. Since blockages closer to the UE have a more pronounced impact on the received signal, we introduce an exponential distance-decay weighting and define the effective contribution of each blocking object as $B_j = L_d(v_j)\,\exp\left(-\frac{d_{O_{\text{LOS},j}-\text{UE}}}{L_{\mathrm{shadow}}}\right)$,
where $L_{\mathrm{shadow}}$ is a scale parameter describing the spatial correlation of shadow fading. Considering that multiple blockages may occur simultaneously, the overall blockage intensity is characterized by the maximum of all $B_j$. To prevent excessively large individual diffraction values from causing over-sensitive blockage estimation, we further introduce an upper threshold $B_{\max}$ and define
\begin{equation}
\small
    \pmb{e}_{\mathrm{OF}}
    = \exp\!\left(-\,\frac{\min\left(\max_j B_j,\; B_{\max}\right)}{K_{\mathrm{dB}}}\right),
\end{equation}
where $K_{\mathrm{dB}}$ maps the blockage intensity in dB to the interval $[0,1]$. This factor has a clear physical interpretation: when the path is unblocked, $\max_j B_j \approx 0$, yielding $\pmb{e}_{\mathrm{OF}} \approx 1$; as buildings increasingly intrude into the first Fresnel zone or their number grows, $\max_j B_j$ increases and $\pmb{e}_{\mathrm{OF}}$ decreases monotonically toward 0.

\subsection{Multipath Propagation Modelling}
\label{sec::multipath}
The urban multipath propagation can be interpreted as a structured physical process composed of LOS propagation and multi-order NLOS propagation. As shown in Fig.~\ref{Fig_syst}, radio waves are influenced by reflections generated on large-scale surfaces such as building facades and the ground. To model signal reflections induced by urban structures and construct a physical representation of multipath effects, we build a multi-order reflection geometric model and map each feasible reflected path into a structured vectorized feature. Let the 3D positions of the BS and UE be denoted by $\pmb{p}_0$ and $\pmb{p}_{n+1}$, respectively. Suppose the signal undergoes $n > 0$ consecutive reflections during a reflected path, the corresponding reflection points can be denoted by $\pmb{p}_1,\ldots,\pmb{p}_{n}$. According to the law of planar reflection, for the $i$-th reflection, the incident angle $\theta_i^{\mathrm{inc}}$ equals the reflected angle $\theta_i^{\mathrm{ref}}$, and both lie in the plane of incidence that is perpendicular to the reflecting surface. Under geometric feasibility and non-occlusion constraints, the series $\{\pmb{p}_0,\pmb{p}_1,\ldots,\pmb{p}_{n+1}\}$ forms a valid $n$-th order reflection path, whose total propagation distance is
\begin{equation}
\small
    d_{\mathrm{ref}}
    = \sum_{i=0}^{n}\left\|\pmb{p}_{i+1}-\pmb{p}_{i}\right\|_2.
    \label{eq:prop_dis}
\end{equation}

Based on reflection geometry, we define the cosine of the departure angle $\cos\theta_{\mathrm{AoD},i}\ (i \geq 0)$ as the directional cosine of the ray propagating from $\pmb{p}_i$ toward the next reflection point $\pmb{p}_{i+1}$, and the cosine of the incident angle $\cos\theta_{\mathrm{inc},i}\ (i \geq 1)$ as the cosine of the angle between the incident direction and the surface normal at the reflection point $\pmb{p}_{i}$. Accordingly, this reflected path can be represented by the following $(2n+3)$-dimensional feature vector:
\begin{equation}
\small
    \pmb{f}_{\mathrm{ref}} =
    \big[
        \log_{10}d_{\mathrm{ref}},\;
        n,\;
        \cos\theta_{\mathrm{AoD},0},\;
        \cos\theta_{\mathrm{inc},1}, \;
        \cos\theta_{\mathrm{AoD},1}, \;
        ... \;
    \big]^{\mathsf{T}},
    \label{eq:ref1}
\end{equation}

Typically, for each position along $\pmb{\Gamma}$, there are multiple reflected paths between the BS and UE. Considering up to $N_{\mathrm{ref}}$ reflections, we define an $(N_{\mathrm{NLOS}}+1) \times (2N_{\mathrm{ref}}+3)$-dimensional reflection embedding $\pmb{e}_\text{RE}$ to represent the LOS path and the top $N_{\mathrm{NLOS}}$ dominant reflected paths:
\begin{equation}
\small
    \pmb{e}_\text{RE}^{(N_{\mathrm{NLOS}}+1) \times (2N_{\mathrm{ref}}+3)} =[\pmb{f}_0^{\mathsf{T}},\,\pmb{f}_1^{\mathsf{T}},\,\ldots,\,\pmb{f}_{N_{\mathrm{NLOS}}}^{\mathsf{T}}]^{\mathsf{T}},
    \label{eq:ref2}
\end{equation}
where $\pmb{f}_0$ corresponds to the LOS path and $\pmb{f}_1,\ldots,\pmb{f}_{N_{\mathrm{NLOS}}}$ correspond to reflected paths. Since reflection losses are typically much larger than path-loss attenuation, lower-order reflections are prioritized; within the same reflection order, paths are sorted from shorter to longer. For paths with fewer than $N_{\mathrm{ref}}$ actual reflections, the remaining entries are padded with zeros. If the number of candidate paths is insufficient, zero vectors are padded to maintain consistent feature dimensionality. The embedding $\pmb{e}_\text{RE}$ preserves geometric interpretability while transforming the complex reflection propagation process into a unified representation suitable for machine learning models.

\section{Physics-informed Conditional Diffusion Models}
\label{sec:PIDM}
In this section, we introduce the physics-informed conditional diffusion models we designed to construct the prediction layer of Channel-Diff. RSRP determination involves multi-scale characteristics influenced by network configuration and the spatial environment, including LS attenuation such as long-distance path loss, blockage, and shadow fading, as well as SS attenuation induced by multipath effects. Due to the influence of spatial environments, medium characteristics, and reflective surface materials, RSRP contains complex spatio-temporal features that are difficult for conventional deep learning predictors to capture. Its relationship with environmental conditions is constrained by signal propagation physics while also exhibiting randomness. Moreover, since the RSRP series is aligned with the user trajectory, it lacks periodic patterns and presents highly complex and stochastic behaviour. Generative models such as GANs often suffer from mode collapse in such scenarios~\cite{10.5555/3540261.3540933}, making it difficult to balance diversity and accuracy. Diffusion models, by contrast, learn data distributions through iterative denoising of random perturbations, making them particularly well suited to capture random fluctuations caused by SS fading.

\begin{figure*}[tb]
    \centering
    \includegraphics[width=\linewidth]{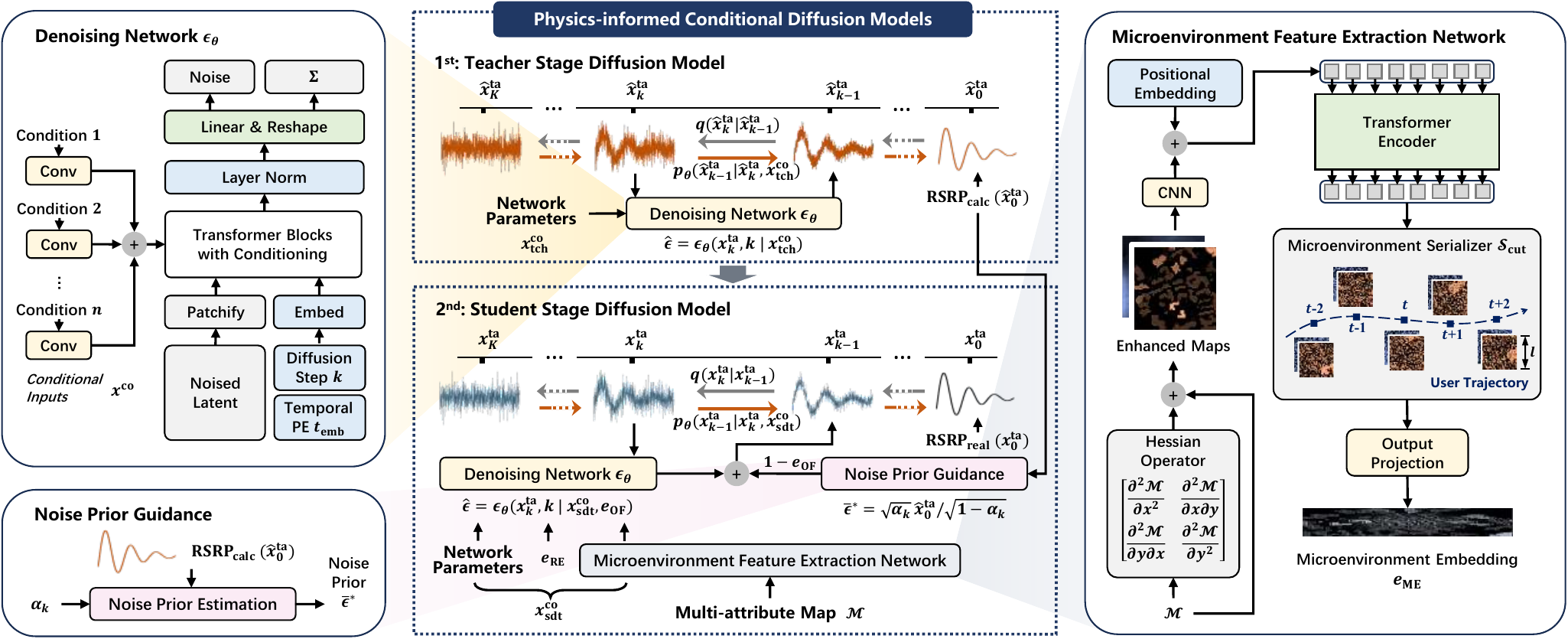}
    \caption{The structure of the physics-informed conditional diffusion models, with the pipeline of the denoising network, noise prior guidance, and the Microenvironment Feature Extraction Network.}
    \label{Fig_PIDM}
\end{figure*}

The physics representation layer has provided effective representations of LS and SS physical processes. To enable the diffusion models to capture the general principles underlying these processes, we custom-design physics-informed conditional diffusion models $\mathcal{P}$. This model maps multimodal information characterizing LS and SS attenuation to RSRP and introduces a two-stage training paradigm for physical-information fusion to enhance interpretability. The structure of the physics-informed conditional diffusion models is shown in Fig.~\ref{Fig_PIDM}.

\subsection{Denoising Network}
The denoising network $\pmb{\epsilon}_\theta$ is a specially designed neural network for noise estimation in diffusion models~\cite{DDPM, CSDI}. According to Eq.~(\ref{eq:reverse_process}), for diffusion step $k$ in the reverse process, the denoising network takes the current noised latent variable $\pmb{x}_k^{\text{ta}}$ and the timestep index $k$ as input, and output the estimated expectation $\mu_\theta$ and variance $\Sigma_\theta$ of the noise. For conditional diffusion models, the denoising network would take an additional conditional input $\pmb{x}^\text{co}$ to estimate a conditional distribution of noise, so that achieving the mapping $\pmb{x}_{k-1}^{\text{ta}} \overset{k,~\pmb{x}^{\text{co}}}{\mapsto} \pmb{x}_{k}^{\text{ta}}$. This process can be denoted as:
\begin{equation}
    \hat{\pmb\epsilon} = \mathcal{N}(\mu_\theta,~\Sigma_\theta\pmb{I})=\pmb{\epsilon}_\theta(\pmb{x}_k^\text{ta}, k|\pmb{x}^\text{co}),
\end{equation}
and $\pmb{\epsilon}_\theta$ is typically trained by Eq.~(\ref{eq:objective}) in the forward process.

In Channel-Diff, we customize the denoising network to achieve multimodal conditioning. Inspired by the adaptive Layer Norm (adaLN) in DiT~\cite{10377858},  we use Transformer blocks with adaLN conditioning scheme to conduct conditional noise distribution estimation. As shown in Fig.~\ref{Fig_PIDM}, the Transformer blocks can take multiple external conditions as input. Each condition is pre-mapped through a convolutional layer, such that raw conditions of different dimensionalities are uniformly projected into the conditional space $\mathcal{X}^\text{co}$. This allows the denoising equation to approximate the conditional distribution of the data under specific conditions during each single-step denoising operation. Further, to enable the neural network to capture temporal positional relationships within the RSRP series, we employ a sinusoidal temporal positional encoding:
\begin{equation}
\small
    \text{pe}^{(t)} =\left[\sin\left({t}/{10^{4·\frac{2i}{D_\text{PE}}}} \right),\,\cos\left({t}/{10^{4·\frac{2i+1}{D_\text{PE}}}} \right)  \right]_{i=0}^{D_\text{PE}/2 - 1},
\end{equation}
\begin{equation}
\small
    \pmb{t}_\text{emb} =\left[\text{pe}^{(0)}, \text{pe}^{(1)}, ...,\text{pe}^{(T)}\right]^{\mathsf{T}},
\end{equation}
where $\pmb{t}_\text{emb} \in \mathbb{R}^{T \times D_\text{PE}}$, $D_\text{PE}$ denotes the total dimension of the positional encoding.

\subsection{Microenvironment Feature Extraction Network}
\label{subsec_MFEN}
Ignoring Doppler effects from UE movement, SS attenuation mainly results from the multipath effect, which is caused by signal reflections and scattering off buildings, the ground, and other urban structures~\cite{PESCE2023441, LEE200246-1}. Modelling the multipath effect in urban spaces typically requires ray tracing simulations~\cite{4061541,10069838,9811058}, which incurs significant time and computational resource overhead. Thus, a fast ML-based method for extracting urban multipath features is needed. In Section~\ref{sec::multipath}, we define a reflection embedding $\pmb{e}_\text{RE}$ to represent the multipath propagation based on macroscopic geometric modelling. However, this modelling may be biased, as we adopt a highly simplified planar assumption for building surfaces and consider only the $N_\text{NLOS}$ dominant reflected paths. The urban microenvironment surrounding the user has a significant impact on multipath effects between the UE and the BS. 2-dimensional features such as high building density and narrow streets forming “urban canyons” are difficult to be fully captured by $\pmb{e}_\text{RE}$. 

To construct an urban microenvironment representation, we propose a Microenvironment Feature Extraction Network (MFEN), as shown in Fig.~\ref{Fig_PIDM}. It is a pipeline-style processing framework for the multi-attribute map $\mathcal{M}$. The main structure of MFEN consists of a Hessian operator, positional embedding, a Transformer-based spatial encoder, and a microenvironment serializer. The Hessian operator is a second-order image differential operator capable of extracting edge features from images. We employ a residual structure to enhance edge features in $\mathcal{M}$, since signal reflections are mainly influenced by building facades. The Hessian matrix is represented as:
\begin{equation}
\small
    \pmb{H}(x, y) = 
    \begin{bmatrix}
        \frac{\partial^2 \mathcal{M}}{\partial x^2} & \frac{\partial^2 \mathcal{M}}{\partial x \partial y} \\
        \frac{\partial^2 \mathcal{M}}{\partial y \partial x} & \frac{\partial^2 \mathcal{M}}{\partial y^2}
    \end{bmatrix}.
\end{equation}

We then use a convolutional layer and a Transformer-based spatial encoder to extract spatial features and encode the images into the embedding space $\mathbb{M}$. To compensate for the lack of position awareness in the CNN, the positional encoding $\pmb{P} \in \mathbb{R}^{2 \times H \times W}$ is used to induce the image with positional labels. $H$ and $W$ are the pixel width and length of the enhanced map. The Transformer-based spatial encoder uses the self-attention mechanism achieved by the transformer encoder to capture spatial dependencies in the image, thereby enhancing the model's understanding of spatial information in the map. The input of the Transformer encoder is represented as $\mathcal{M}^*=Conv(\mathcal{M}+\pmb{H}\mathcal{M})+\pmb{P}$, and the output can be denoted as:
\begin{equation}
\small
    m = \text{softmax} \left( \frac{m_Qm_K^T}{\sqrt{d_\text{attn}}} \right) m_V \in \mathbb{M},
\end{equation}
where $m_Q=\mathcal{M}^* \times W^Q$, $m_K=\mathcal{M}^* \times W^K$ and $m_V=\mathcal{M}^* \times W^V$. $W^Q, W^K, W^V$ are learnable weight matrices, $d_\text{attn}$ is the attention scaling factor.

To align the environment representation with the user's movement trajectory in both space and time, we employ a microenvironment serializer $\mathcal{S}_{\text{cut}}$. It converts the static encoded map $m$ into a micro-map series. Given the Field of View (FoV) $l$ measured in meter, $\mathcal{S}_{\text{cut}}$ crops a $l \times l~\text{m}^2$ sub-region around each point on the user trajectory $\pmb{\Gamma}$, producing a $T$-length micro-map series $\pmb{m}$. This process can be formulated as
\begin{equation}
\small
    \pmb{m} = \mathcal{S}_{\text{cut}}(m, \pmb\Gamma, l),
\end{equation}
Finally, we use an MLP to project the micro-map series $\pmb{m}$ into a temporal microenvironment embedding $\pmb{e}_\text{ME}=\text{MLP}(\pmb{m})$. The microenvironment embedding $\pmb{e}_\text{ME}$ represents regional microenvironment features aligned with the user trajectory, which would highly influence the signal's SS attenuation.

\subsection{Prior Knowledge Guided Two-stage Diffusion Models}
\label{sec::two-stage}
The prior knowledge guided two-stage training scheme is the core architecture of the proposed Physics-informed Conditional Diffusion Models. Inspired by the teacher–student paradigm in knowledge distillation, we divide the training of the diffusion model into a teacher stage and a student stage, enabling the model to learn the electromagnetic physical principles underlying multi-scale signal attenuation. As shown in Fig.~\ref{Fig_syst}, the input layer and the physics representation layer provide the prediction layer with multimodal conditional inputs and physical representations that describe multi-scale channel states. The middle part of Fig.~\ref{Fig_syst} presents a detailed illustration of the architecture of the two-stage training of diffusion models.

\subsubsection{Teacher Stage}
The network parameters serve as common inputs for both the teacher and student training stages. In the teacher stage, the training label is $\text{RSRP}_\text{calc}$. The model can be trained efficiently to fit measurement-based propagation models. The model initialized through the teacher stage can thus be regarded as having learned the physical principles underlying LS LOS propagation. Thus, in the teacher stage, the conditional input of the denoising network can be represented as $\pmb{x}^\text{co}_\text{tch}=[\text{Network Parameters}]$. The total loss in the teacher stage is the Mean Squared Error (MSE) between the generated vector $\pmb{x}^{\text{ta}}_k$ and the theoretical target vector $\hat{\pmb{x}}^{\text{ta}}_0 = \textbf{RSRP}_\text{calc}$, expressed as
\begin{equation}
\small
\mathcal{L}_{\text{teacher}} = \text{MSE}(\hat{\pmb{x}}^{\text{ta}}_0, \pmb{x}^{\text{ta}}_k(\pmb{x}^{\text{ta}}_K | \pmb{x}^{\text{co}}_{\text{tch}})),
\end{equation}
where $\pmb{x}^{\text{ta}}_K$ is the initial value of the reverse process, i.e., random noise, and $K$ is the total number of denoising steps.

\subsubsection{Student Stage with Noise Prior Guidance}
Compared with the teacher stage, the student stage introduces more complex LS and SS features that are difficult to describe analytically, including $\pmb{e}_\text{RE}$, $\pmb{e}_\text{OF}$, and the multi-attribute map $\mathcal{M}$. To fully inherit the physical knowledge of LS LOS propagation learned by the teacher-stage model, we design an occlusion-factor-weighted Noise Prior Guidance mechanism. According to Section~\ref{sec::CDM}, the forward process can be formulated as $\pmb{x}^{\text{ta}}_k = \sqrt{\alpha_k}\pmb{x}^{\text{ta}}_0 + (1-\alpha_k)\pmb\epsilon$. Assume that $\pmb{x}^{\text{ta}}_0$ is influenced by multiple factors and can be decomposed into a weighted sum of a deterministic component $\pmb{X}^{\text{ta}}$ and a stochastic component $\Delta \pmb{x}^{\text{ta}}_0$, i.e., $\pmb{x}^{\text{ta}}_0 = e \cdot \pmb{X}^{\text{ta}} + (1-e) \cdot \Delta \pmb{x}^{\text{ta}}_0$, where $e \in [0,1]$ is a weighting factor that measures the reliability of $\pmb{X}^{\text{ta}}$ as an estimate of $\pmb{x}^{\text{ta}}_0$. Therefore, the forward process can be expressed as
\begin{equation}
    \pmb{x}^{\text{ta}}_k = \sqrt{\alpha_k}\left[e \cdot \pmb X^{\text{ta}}+(1-e) \cdot \Delta \pmb x^{\text{ta}}_0\right] + (1-\alpha_k)\pmb\epsilon.
\end{equation}
This expression can be equivalently rewritten as
\begin{equation}
\small
    \begin{aligned}
    \boldsymbol{\epsilon} =& \frac{\pmb{x}^{\text{ta}}_0 - \sqrt{\alpha_k} (e\pmb X^{\text{ta}} + (1-e)\Delta\pmb{x}^{\text{ta}}_0)}{{1 - \alpha_k}} \\
    =& \frac{\pmb{x}^{\text{ta}}_0 - \sqrt{\alpha_k} e\Delta\pmb{x}^{\text{ta}}_0}{{1 - \alpha_k}}
    - \frac{\sqrt{\alpha_k} (1-e)\pmb X^{\text{ta}}}{{1 - \alpha_k}}.
    \end{aligned}
    \label{eq_prior}
\end{equation}
Define residual noise $\Delta \pmb \epsilon=\frac{\pmb{x}^{\text{ta}}_0 - \sqrt{\alpha_k} e\Delta\pmb{x}^{\text{ta}}_0}{{1 - \alpha_k}}$, and prior noise $\bar{\pmb \epsilon}=\frac{\sqrt{\alpha_k} (1-e)\pmb X^{\text{ta}}}{{1 - \alpha_k}}=(1-e)\cdot\frac{\sqrt{\alpha_k} \pmb X^{\text{ta}}}{{1 - \alpha_k}}:=(1-e)\cdot{\bar{\pmb \epsilon}}^*$, the noise $\boldsymbol{\epsilon}$ can be denoted by $\boldsymbol{\epsilon}=\Delta \pmb \epsilon-\bar{\pmb \epsilon} \Leftrightarrow \Delta \pmb \epsilon=\boldsymbol{\epsilon}+(1-e){\bar{\pmb \epsilon}}^*$.

In Channel-Diff, the occlusion factor $\pmb{e}_{\text{OF}}$ quantifies the reliability of $\textbf{RSRP}_\text{calc}~(:=\hat{\pmb{x}}^{\text{ta}}_0)$. In LOS scenarios, propagation models can typically provide more accurate PL estimation. Accordingly, we construct the Noise Prior Guidance mechanism for the student-stage diffusion model using $\textbf{RSRP}_\text{calc}$, i.e.,
\begin{equation}
   {\bar{\pmb \epsilon}}^*=\frac{\sqrt{\alpha_k} \hat{\pmb{x}}^{\text{ta}}_0}{{1 - \alpha_k}},~~~\pmb \epsilon=\Delta \pmb \epsilon+(1-\pmb{e}_{\text{OF}})\cdot{\bar{\pmb \epsilon}}^*.
\end{equation}
Moreover, in the forward process of the student stage, $\Delta \pmb{\epsilon} = \frac{\pmb{x}^{\text{ta}}_0 - \sqrt{\alpha_k}\,\pmb{e}_{\text{OF}}\,\Delta \pmb{x}^{\text{ta}}_0}{1 - \alpha_k}$ is correlated with $\pmb{e}_{\text{OF}}$. Therefore, the denoising network should be formulated as
\begin{equation}
    \hat{\Delta \pmb \epsilon}=\pmb{\epsilon}_\theta(\pmb{x}_k^\text{ta},k|\pmb{x}_\text{sdt}^\text{co},\pmb{e}_{\text{OF}}),
\end{equation}
where $\pmb{x}_\text{sdt}^\text{co}=[\text{Network Parameters},~\pmb{e}_\text{RE},~\pmb{e}_\text{ME}]$. We retrain the model pre-trained in the teacher stage using real RSRP data $\pmb{x}^{\text{ta}}_0=\textbf{RSRP}_\text{real}$, enabling it to further learn SS attenuation characteristics induced by multipath effects. The reflection embedding $\pmb{e}_\text{RE}$ provides a macroscopic representation of multipath propagation, while the multi-attribute map $\mathcal{M}$ is processed by the MFEN to obtain the microenvironment embedding $\pmb{e}_\text{ME}$, which serves as a microscopic representation of multipath propagation. The loss function for student stage training is:
\begin{equation}
\small
    \mathcal{L}_{\text{student}} = \text{MSE}(\pmb{x}^{\text{ta}}_0, \pmb{x}^{\text{ta}}_k(\pmb{x}^{\text{ta}}_K | \pmb{x}^{\text{co}}_{\text{sdt}},\pmb{e}_{\text{OF}})).
\label{eq:loss_stu}
\end{equation}


\section{Evaluation}
\label{sec:Eval}
We use two publicly available real-measured RSRP datasets to evaluate the performance of Channel-Diff and answer the following Research Questions (RQ):
\begin{itemize}
    \item \textbf{\textit{RQ1}}: How does Channel-Diff perform on the RSRP prediction task compared with existing models?
    \item \textbf{\textit{RQ2}}: Does the physics-guided two-stage diffusion training strategy improve RSRP prediction performance and model interpretability?
    \item \textbf{\textit{RQ3}}: What performance gains do SS embeddings, including reflection embeddings and urban microenvironment embeddings, provide for RSRP prediction?
\end{itemize}

\subsection{Dataset}
\subsubsection{\textbf{RSRP-CPGMCM}}
This is a real-measured RSRP dataset collected by 8 terminal devices, containing 5,259,757 points with RSRP and corresponding factors. We organize this dataset into 80,000 RSRP sequences of length $T = 64$, and select 10-dimensional LS factors required by Channel-Diff from the given factors. This dataset does not contain the map data necessary for the multi-attribute map. Given the scarcity of real-world RSRP datasets, we can still use this dataset to validate Channel-Diff's capability in generating RSRP guided by physical knowledge. For more details, please refer to~\cite{RSRP-CPGMCM}.

\subsubsection{\textbf{RSRP-Image}}
This is a real-measured radio map dataset. It contains 180 radio maps covering 320m × 320m regions, along with the corresponding building height maps, ground elevation maps, and BS parameters. The radio maps are generated by interpolating discrete RSRP measurements using the K-means method. Based on the building distributions, we randomly generated 20 user trajectories within each map, resulting in a total of 3,600 sequence samples. Using the provided information, we choose and calculate 10 LS factors corresponding to each trajectory. For more details, please refer to~\cite{RSRP-Image}.

\subsection{Baselines}
We select 13 representative baseline models to verify the performance superiority of Channel-Diff. These models include statistical models: PM~\cite{Hata, WINNERII} and Multiple Linear Regression (MLR)~\cite{Behjati2022}); special ML-based models: WPM-BPNN~\cite{BPNN}, LSTM-LQE~\cite{LSTM-10575638}, GBDT+DNN~\cite{GBDT-9700950}, VAE+DNN~\cite{VAE-9978065}) and EFEM~\cite{9351998}; radio map generation models: RadioGAN~\cite{EM-10227351} and RadioDiff~\cite{EM-2024arXiv240808593W}; GANs-based multivariate time series generation model: MTC-CGAN~\cite{MTS-CGAN-2022arXiv221002089M} and TimeGAN~\cite{10.5555/3454287.3454781}; as well as diffusion-models-based generation models: CSDI~\cite{CSDI} and CANDLE~\cite{10.1145/3678717.3691312}. Table \ref{Tab_bs} presents a comparative overview of the selected baselines and our proposed Channel-Diff model, considering four key aspects: the RSRP data format, the use of machine learning, the incorporation of SS features, and the inclusion of physics-informed mechanisms.
\begin{table*}[tb]
\centering
\caption{Comparison of the characteristics of Channel-Diff and various baselines.}
\begin{tabular}{ccccccc}
\toprule
\textbf{Type} & \textbf{Method} & \textbf{RSRP Format} & \textbf{ML-based} & \textbf{SS Feature Considered} & \textbf{Physics-informed} \\
\midrule
\multirow{2}{*}{Statistical Models} 
    & PM~\cite{Hata, WINNERII} & \textit{point} & \ding{55} & \ding{55} & \ding{55} \\
    & MLR~\cite{Behjati2022} & \textit{point} & \ding{55} & \ding{55} & \ding{55} \\
\midrule
\multirow{5}{*}{Special Models} 
    & WPM-BPNN~\cite{BPNN} & \textit{point} & \ding{51} & \ding{55} & \ding{55} \\
    & LSTM-LQE~\cite{LSTM-10575638} & \textit{sequence} & \ding{51} & \ding{55} & \ding{55} \\
    & GBDT+DNN~\cite{GBDT-9700950} & \textit{point} & \ding{51} & \ding{55} & \ding{55} \\
    & VAE+DNN~\cite{VAE-9978065} & \textit{point} & \ding{51}& \ding{51} & \ding{55} \\
    & EFEM~\cite{9351998} & \textit{point} & \ding{51}& \ding{51} & \ding{55} \\
\midrule
\multirow{2}{*}{Radio Map} 
    & RadioGAN~\cite{EM-10227351} & \textit{image} & \ding{51} & \ding{51} & \ding{55} \\
    & RadioDiff~\cite{EM-2024arXiv240808593W} & \textit{image} & \ding{51} & \ding{51} & \ding{55} \\
\midrule
\multirow{2}{*}{GANs}
    & MTS-CGAN~\cite{MTS-CGAN-2022arXiv221002089M} & \textit{sequence} & \ding{51} & \ding{55} & \ding{55} \\
    & TimeGAN~\cite{10.5555/3454287.3454781} & \textit{sequence} & \ding{51} & \ding{55} & \ding{55} \\
\midrule
\multirow{3}{*}{Diffusion Models}
    & CSDI~\cite{CSDI} & \textit{sequence} & \ding{51} & \ding{55} & \ding{55} \\
    & CANDLE~\cite{10.1145/3678717.3691312} & \textit{sequence} & \ding{51} & \ding{55} & \ding{55} \\
    & Channel-Diff (Ours) & \textit{sequence} & \ding{51} & \ding{51} & \ding{51} \\   
\bottomrule
\end{tabular}
\begin{center}
    Hint: \textbf{RSRP Format} refers to the structure of the RSRP data that the model processes directly. \ding{51} under \textbf{ML based} means the method is based on ML. \ding{51} under \textbf{SS Feature Considered} means the method considers the SS factors extracted from the urban environment. \ding{51} under \textbf{Physics-informed} means the method utilizes the physical knowledge of signal propagation.
\end{center}
\label{Tab_bs}
\vspace{-4mm}
\end{table*}

\subsection{Evaluation Metrics}
\label{subsec::metrics}

\subsubsection{Jensen-Shannon Divergence (JSD)}
A symmetric and smoothed measure of divergence between two probability distributions. Given two distributions $p$ and $q$, the JSD is defined as:
\begin{equation}
\small
    \mathrm{JSD}(p \,\|\, q) = \frac{1}{2} \mathrm{KL}(p \,\|\, M) + \frac{1}{2} \mathrm{KL}(q \,\|\, M),
\end{equation}
where $M = \frac{1}{2}(p + q)$ is the average distribution, and $\mathrm{KL}(p \,\|\, q)$ denotes the Kullback-Leibler divergence.

\subsubsection{Normalized Root Mean Squared Error (NRMSE)}
The Root Mean Squared Error (RMSE) normalized by the range or mean of the true values, providing a scale-independent measure of error:
\begin{equation}
\small
    \mathrm{NRMSE} = \frac{\sqrt{\frac{1}{n} \sum_{i=1}^n (y_i - \hat{y}_i)^2}}{y_{\text{max}} - y_{\text{min}}},
\end{equation}
where $y_i$ and $\hat{y}_i$ are the true and predicted values, respectively. A lower NRMSE indicates better predictive performance, with values closer to 0 representing higher accuracy.

\subsubsection{Mean Absolute Error (MAE)}
The average magnitude of the errors between predicted and true values, without considering their direction:
\begin{equation}
\small
    \mathrm{MAE} = \frac{1}{n} \sum_{i=1}^n \left| y_i - \hat{y}_i \right|,
\end{equation}
where $y_i$ is the true value and $\hat{y}_i$ is the predicted value.

\subsection{Overall Performance (\textbf{\textit{RQ1}})}
\label{main_perf}
In this section, we evaluate the RSRP prediction performance of Channel-Diff on RSRP-CPGMCM and RSRP-Image. Table~\ref{Tab_main} presents the evaluation results of Channel-Diff compared with 13 baselines. Here, $\Delta$ denotes the percentage improvement of Channel-Diff over a given model for each metric, while $\bar{\Delta}$ represents the average improvement across all three metrics, reflecting the overall gain in both distributional and absolute prediction accuracy. As shown in Table~\ref{Tab_main}, Channel-Diff achieves the best overall performance across all baselines on both datasets. On the RSRP-CPGMCM dataset, Channel-Diff improves upon the second-best model, GBDT+DNN, by 37.19\% on average; on the RSRP-Image dataset, it outperforms the second-best model, EMEF, by 25.15\%. Notably, although the small models specifically designed for RSRP prediction are primarily aimed at single-point RSRP tasks, they still outperform general-purpose time series models in sequence prediction. This highlights the necessity of task-specific architectural design for RSRP prediction. Furthermore, on the RSRP-Image dataset, models such as EMEF and RadioGAN, which incorporate urban environmental information related to SS features, significantly outperform those that do not. 
\begin{table*}[tb]
\centering
\caption{Performance comparison of different methods on RSRP-CPGMCM and RSRP-Image.}
\setlength{\tabcolsep}{3pt}
\begin{tabular}{cccccccccccccccc}
\toprule
\multirow{3}{*}{Method} & \multicolumn{6}{c}{RSRP-CPGMCM} & \multirow{3}{*}{$\bar{\Delta}$} & \multicolumn{6}{c}{RSRP-Image} & \multirow{3}{*}{$\bar{\Delta}$} \\
\cmidrule(lr){2-7} \cmidrule(lr){9-14}
 & JSD$\downarrow$ & $\Delta$ & NRMSE$\downarrow$ & $\Delta$ & MAE$\downarrow$ & $\Delta$ & & JSD$\downarrow$ & $\Delta$ & NRMSE$\downarrow$ & $\Delta$ & MAE$\downarrow$ & $\Delta$ & \\
\midrule
PM
& 0.3867 & 69.20\% & 0.4971 & 59.18\% & 15.27 & 62.80\% & 63.73\%
& \underline{0.1454} & 3.65\% & 0.4869 & 45.06\% & 10.73 & 48.56\% & 32.42\% \\

MLR
& 0.3962 & 69.94\% & 0.2557 & 20.65\% & 7.624 & 25.49\% & 38.69\%
& 0.4387 & 68.06\% & \underline{0.3038} & 11.95\% & \underline{6.962} & 20.71\% & 33.58\% \\

\midrule
WPM-BPNN
& 0.3976 & 70.05\% & \underline{0.2522} & 19.55\% & \underline{7.54} & 24.66\% & 38.08\%
& 0.4166 & 66.37\% & 0.3150 & 15.08\% & 7.415 & 25.56\% & 35.67\% \\

LSTM-LQE
& 0.7547 & 84.22\% & 0.3783 & 46.37\% & 10.19 & 44.25\% & 58.28\%
& 0.7515 & 81.36\% & 0.4364 & 38.70\% & 8.968 & 38.45\% & 52.84\% \\

GBDT+DNN
& \underline{0.3225} & 63.07\% & 0.2606 & 22.14\% & 7.714 & 26.35\% & \underline{37.19\%}
& 0.3255 & 56.96\% & 0.3350 & 20.15\% & 7.473 & 26.13\% & 34.41\% \\

VAE+DNN
& 0.3284 & 63.73\% & 0.2728 & 25.62\% & 8.15 & 30.29\% & 39.88\%
& 0.2571 & 45.51\% & 0.3438 & 22.19\% & 7.718 & 28.48\% & 32.06\% \\

EFEM
& - & - & - & - & - & - & -
& 0.1739 & 19.44\% & 0.3605 & 25.80\% & 7.910 & 30.21\% & \underline{25.15\%} \\

\midrule
RadioGAN
& - & - & - & - & - & - & -
& 0.2517 & 44.34\% & 0.3407 & 21.49\% & 7.539 & 26.78\% & 30.87\% \\

RadioDiff
& - & - & - & - & - & - & -
& 0.2209 & 36.58\% & 0.5512 & 51.47\% & 11.56 & 52.25\% & 46.77\% \\

\midrule
MTS-CGAN
& 0.4345 & 72.59\% & 0.3419 & 40.66\% & 9.519 & 40.32\% & 51.19\%
& 0.4153 & 66.27\% & 0.3723 & 28.15\% & 8.191 & 32.61\% & 42.34\% \\

TimeGAN
& 0.6183 & 80.74\% & 0.2781 & 27.04\% & 8.183 & 30.58\% & 46.12\%
& 0.7462 & 81.22\% & 13.79 & 98.06\% & 8.230 & 32.93\% & 70.74\% \\

\midrule
CSDI
& 0.5219 & 77.18\% & 0.5025 & 59.62\% & 15.29 & 62.84\% & 66.55\%
& 0.7135 & 80.36\% & 0.9094 & 70.59\% & 21.7 & 74.56\% & 75.17\% \\

CANDLE
& 0.3362 & 64.57\% & 0.3136 & 35.30\% & 9.734 & 41.64\% & 47.17\%
& 0.3219 & 56.48\% & 0.4376 & 38.87\% & 8.840 & 37.56\% & 44.30\% \\

\midrule
\textbf{Channel-Diff}
& \textbf{0.1191} & - & \textbf{0.2029} & - & \textbf{5.681} & - & -
& \textbf{0.1401} & - & \textbf{0.2675} & - & \textbf{5.520} & - & - \\
\bottomrule
\end{tabular}
\begin{center}
    Hint: $\downarrow$ means lower is better. Bold numbers denote the best results. \underline{Underline} numbers denote the second-best results. $\Delta$ represents the percentage improvement. $\bar\Delta$ represents the average percentage improvement.
\end{center}
\label{Tab_main}
\end{table*}

\begin{figure*}[tb]
\centering
\includegraphics[width=\linewidth]{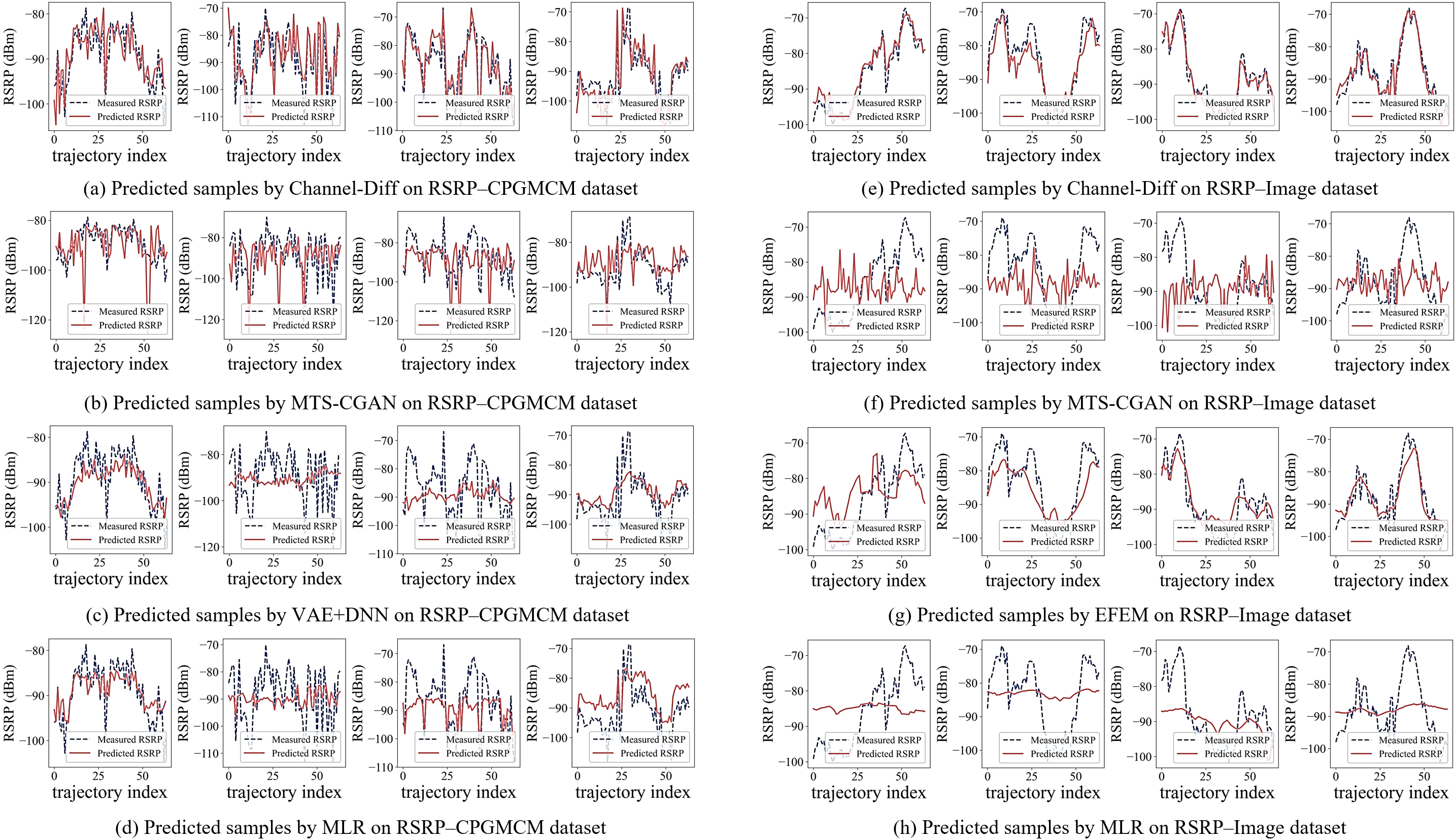}
\caption{Visualization of predicted samples by Channel-Diff and several baselines. (a)–(d) correspond to the RSRP–CPGMCM dataset. (e)–(h) correspond to the RSRP–Image dataset.}
\label{Fig_main}
\vspace{-4mm}
\end{figure*}

Fig.~\ref{Fig_main} presents the visualization of predicted samples by Channel-Diff and several baselines. MTS-CGAN is a GAN-based time-series generation model. The VAE+DNN model constructs a feature extraction network based on VAE and uses DNNs as the output layer. EMEF includes an urban environmental feature extractor built on a VGG-4 network to extract SS attenuation features. MLR, as a non–machine learning model, predicts RSRP by fitting an explicitly defined regression equation; therefore, its input conditions can only incorporate observable and quantifiable network parameters. Comparing (a), (e) with (b), (f), we observe that the GAN-based MTS-CGAN exhibits limitations in fitting the overall trend of the RSRP curve and fails to capture diverse fluctuation patterns. This essentially exposes the inherent mode-collapse problem of GANs. By contrast, although the VAE+DNN model has a simple architecture, it fits the overall trend more effectively. However, it lacks the ability to capture high-frequency fluctuation characteristics. This occurs because the dominant objective of VAE training is to maximize reconstruction of the overall data distribution, causing the decoder to generate averaged and smoothed outputs. EMEF, despite its relatively simple structure built from a VGG-4 layers, achieves better alignment with the overall shape of the RSRP curve thanks to more comprehensive factor consideration. This underscores the importance of including SS conditions that characterize the urban environment in improving RSRP prediction performance.

\begin{figure}[tb]
\centering
\includegraphics[width=\linewidth]{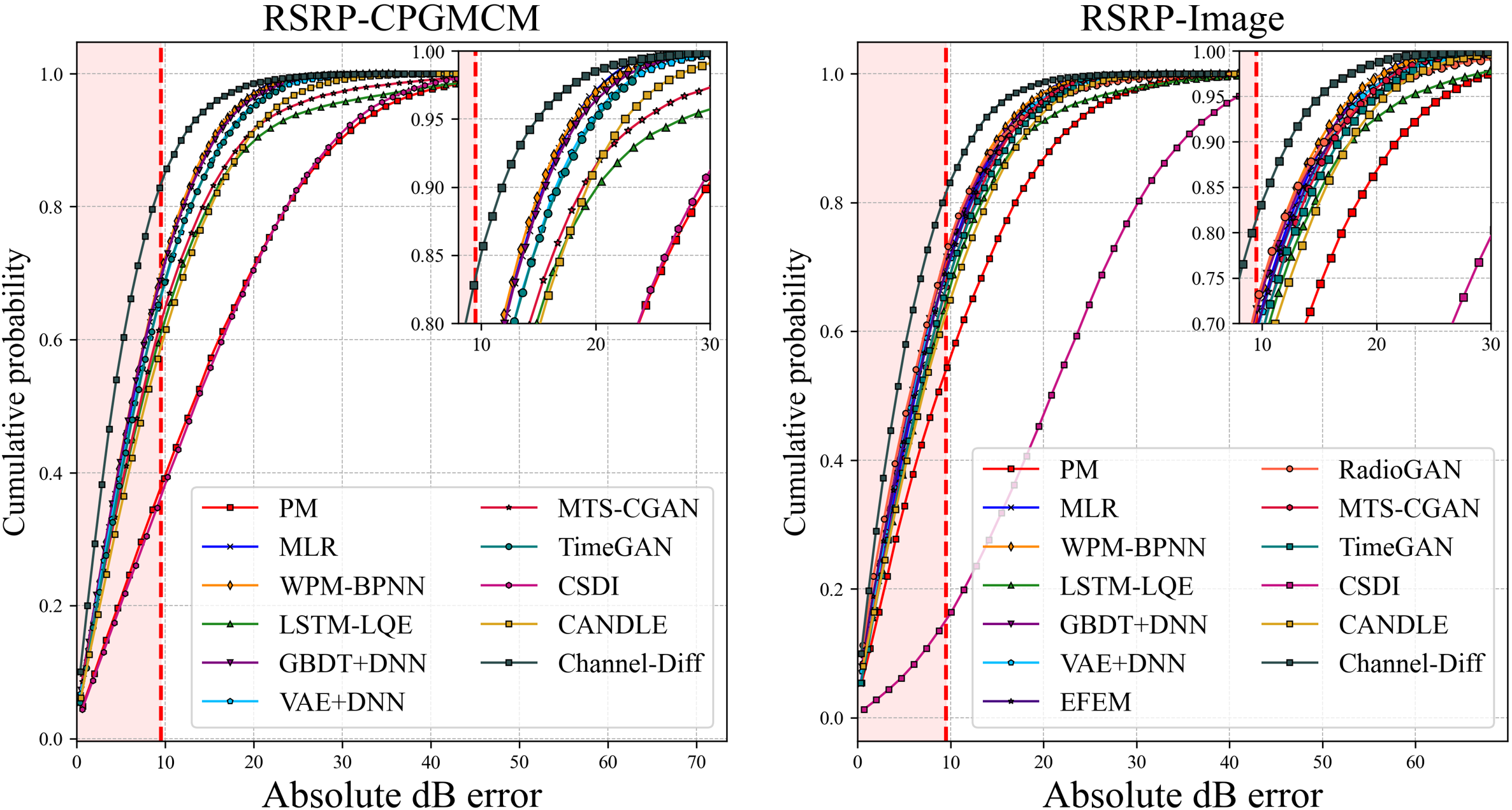}
\caption{Cumulative distribution error curves between real and predicted RSRP by Channel-Diff and baselines on RSRP-CPGMCM and RSRP-Image dataset. The vertical red dashed line indicates the threshold of 9.5 dB, which corresponds to the RSRP measurement accuracy requirement defined in 5G (NR) TS 38.133.}
\label{Fig_cdf}
\vspace{-4mm}
\end{figure}

5G (NR) TS38.133~\cite{3gpp_ts_38_133_v15_3_0} specifies that the measurement accuracy requirement for RSRP under extreme conditions is ±9.5 dB. Fig.~\ref{Fig_cdf} presents the cumulative error distribution curves of Channel-Diff and baseline models on the RSRP-CPGMCM and RSRP-Image datasets, respectively. The vertical red dashed line indicates the threshold of 9.5 dB. It can be observed that Channel-Diff exhibits the lowest overall estimation error distribution, outperforming all baseline methods. Notably, on the RSRP-CPGMCM dataset, approximately 83\% of RSRP predictions have an absolute dB error below 9.5 dB; on the RSRP-Image dataset, this proportion is around 82\%. This level of accuracy is more than sufficiently accurate for RSRP prediction, as 9.5 dB represents the required precision for real-world RSRP measurements rather than prediction. Moreover, the results shown in Fig.~\ref{Fig_cdf} are consistent with those in Table~\ref{Tab_main}, further confirming the outstanding performance of Channel-Diff in the RSRP prediction task. These findings clearly demonstrate the importance of incorporating physical knowledge and SS conditions for accurate RSRP prediction.

To enhance reproducibility, we provide the model parameters and training configurations for the best-performing model on the RSRP-Image dataset in Table~\ref{Tab_main}, as shown in Table~\ref{Tab_parameters}.
\begin{table}[!h]
\centering
\caption{Key Model Parameters and Training Settings.}
\begin{tabular}{p{2cm} p{4.8cm} c}
\toprule
\textbf{Parameter Type} & \textbf{Parameter} & \textbf{Value} \\
\midrule
\multirow{5}{*}{Training}
    & Batch Size & 48 \\
    & Learning Rate & 1e-4 \\
    & Teacher Epochs & 50 \\
    & Student Epochs & 320 \\
    & Early Stop & 12 \\
\midrule
\multirow{6}{*}{Diffusion Models}
    & Diffusion Steps & 400  \\
    & Model Parameters & 66.5M \\
    & Denoising Network: Hidden Size & 256 \\
    & Denoising Network: Attention Heads & 6 \\
    & Denoising Network: Transformer Blocks & 12 \\
    & Condition Channels & 256 \\
\midrule
\multirow{9}{*}{MFEN}
    & FOV & 85m \\
    & $\mathcal{M}$ Dimensions & 2 \\
    & $\mathcal{M}$ Embedding  Dimensions & 256 \\
    & $\pmb{e}_\text{ME}$ Embedding  Dimensions & 256 \\
    & CNN Kernel & 3 \\
    & CNN Layers & 3 \\
    & CNN Hidden Size & 64 \\
    & Attention Heads & 8 \\
    & Attention Model Dimensions  & 256 \\
\midrule
\multirow{3}{*}{Others}
    & Polygon Edges & 3$\sim$6 \\
    & $N_\text{ref}$ & 1 \\
    & $N_\text{NLOS}$ & 4 \\
\bottomrule
\end{tabular}
\label{Tab_parameters}
\end{table}

\subsection{Impact of Prior Knowledge Guidance (\textbf{\textit{RQ2}})}
Channel-Diff incorporates physical knowledge into model parameters through physics representation modules (OSM and MPM) and the prior knowledge guided two-stage diffusion model. 

\subsubsection{Two-stage Training Scheme}
According to Section~\ref{sec::two-stage}, we propose a two-stage training scheme achieve prior knowledge guidance for diffusion models. In the teacher stage, the theoretical RSRP estimated by propagation models is used as the training label, and network parameters serve as the input. Through fast fitting, the model learns the LS LOS electromagnetic propagation knowledge reflected by the propagation models. In the student stage, the pre-trained model from the teacher stage is further trained with real RSRP data. The occlusion-factor-based prior-noise guidance mechanism consolidates the LS physical knowledge learned in the teacher stage, while the reflection embedding and MMFN capture SS attenuation characteristics.

\begin{figure}[tb]
\centering
\includegraphics[width=0.9\linewidth]{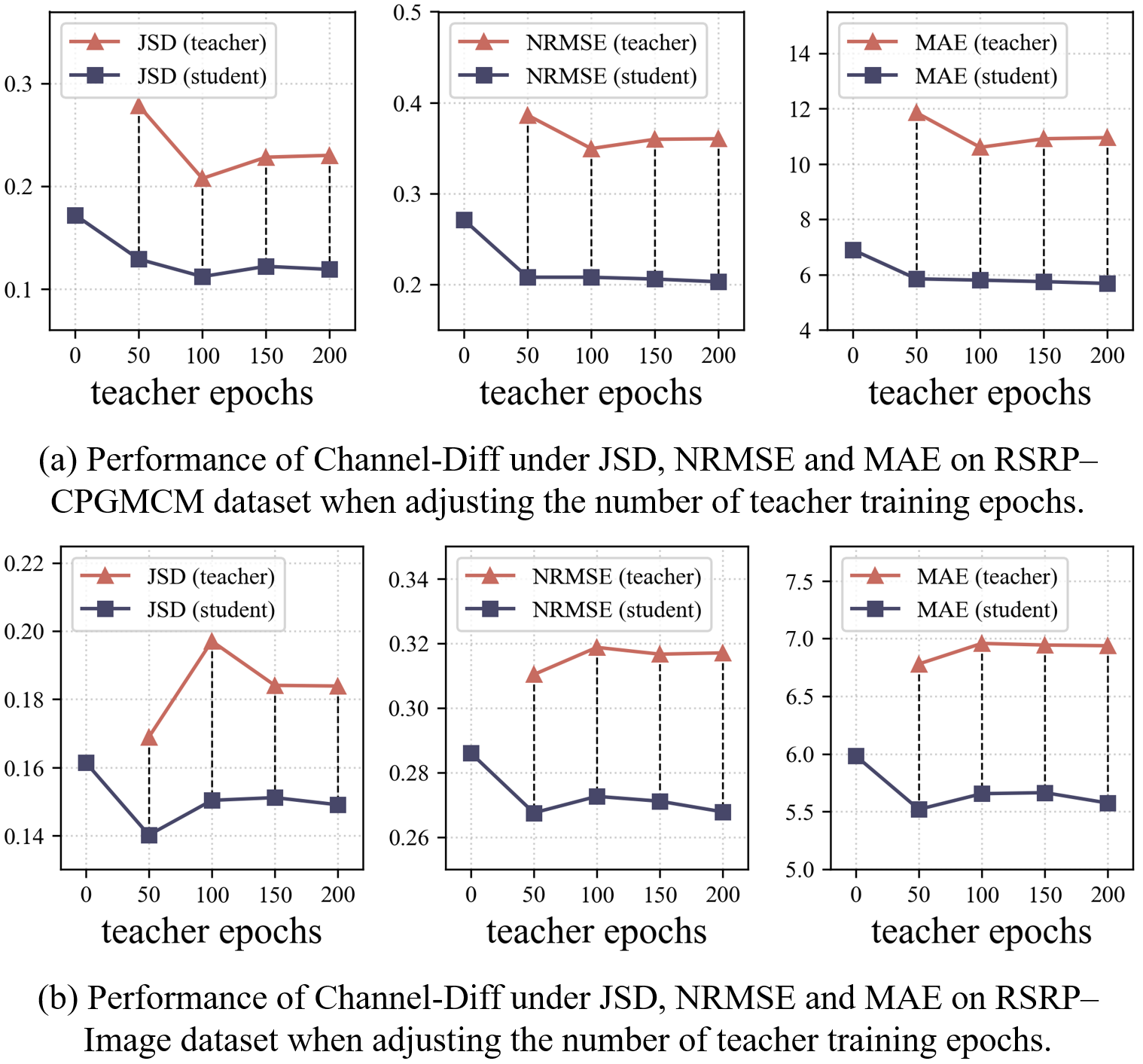}
\caption{Performance of Channel-Diff under JSD, NRMSE, and MAE on RSRP–CPGMCM dataset and RSRP–Image dataset when adjusting the number of teacher training epochs.}
\label{Fig_para}
\vspace{-4mm}
\end{figure}

In this section, we investigate the impact of the number of pre-training epochs in the teacher stage on the model’s prediction performance. Fig.~\ref{Fig_para} shows model performance curves on the RSRP-CPGMCM and RSRP-Image datasets under different numbers of teacher-stage training epochs. The red curve represents the RSRP prediction performance during the teacher stage, while the blue curve represents the prediction performance after completing student-stage training based on the teacher pre-trained model. Because $\textbf{RSRP}_\text{calc}$ is computed using propagation models and has a concise formulaic relationship with network parameters, the teacher-stage training converges rapidly. When the number of epochs is too large, slight overfitting may occur. Due to the discrepancy between $\textbf{RSRP}_\text{calc}$ and $\textbf{RSRP}_\text{real}$, the pre-trained teacher model cannot reach the prediction accuracy of the student-stage model; however, it is evident that the teacher stage leads to a significant improvement in final model performance. Notably, Fig.~\ref{Fig_para} reveals an interesting phenomenon: there is a positive correlation between the final prediction accuracy and the performance of the pre-trained model. Better evaluation metrics of the pre-trained model correspond to better evaluation metrics of the final model. This suggests that the optimal number of teacher-stage epochs lies around 50–100, enabling the model to sufficiently learn prior physical knowledge while effectively avoiding overfitting caused by simple algebraic relationships.

\subsubsection{Noise Prior Gridance}
The occlusion factor $\pmb{e}_\text{OF}$ measures the degree to which buildings in urban space obstruct signal rays. When $\pmb{e}_\text{OF}=1$, i.e., when there is a fully unobstructed LOS condition within the $1^\text{st}$ Fresnel zone between the BS and UE, propagation models can compute RSRP relatively accurately. In the physics-informed conditional diffusion model, we construct a noise prior guidance mechanism based on $\pmb{e}_\text{OF}$, allowing prior physical knowledge of LS LOS propagation to provide stronger guidance for RSRP prediction under unblocked and weakly blocked conditions. We conduct ablation experiments on the noise prior guidance mechanism to verify this hypothesis. To assess the relative overall performance of each model, we introduce the metric $R_{\text{AVG}}$, defined as the sum of a model’s rankings across all three evaluation metrics (JSD, NRMSE, and MAE):
\begin{equation}
\small
    R_{\text{AVG}}=\text{rank(}\text{JSD})+\text{rank(}\text{NRMSE})+\text{rank(}\text{MAE}).
\end{equation}
A lower $R_{\text{AVG}}$ indicates better overall performance. 

Fig.~\ref{Fig_ablation} presents the results of all ablation experiments. Comparing the first and third major columns (the performance metrics of the model without noise prior guidance), we observe that incorporating noise prior guidance yields a significant positive gain in RSRP prediction accuracy. This confirms that the LS LOS prior knowledge embedded in $\textbf{RSRP}_\text{calc}$ provides positive guidance for RSRP prediction during the student-stage training. This mechanism effectively prevents the retrained model from forgetting prior knowledge.

\subsection{Impact of the Small-scale Attenuation Modelling (\textbf{\textit{RQ3}})}
In this section, we conduct a set of experiments to verify whether Channel-Diff’s modelling and feature extraction of SS attenuation lead to effective gains in RSRP prediction. This includes analysing how the modules within MFEN influence the extraction and learning of SS microenvironment features, as well as how efficient multipath propagation modelling yields an effective reflection embedding $\pmb{e}_\text{RE}$.

\subsubsection{Microenvironment Feature Extraction Network}
\begin{figure}[tb]
\centering
\includegraphics[width=\linewidth]{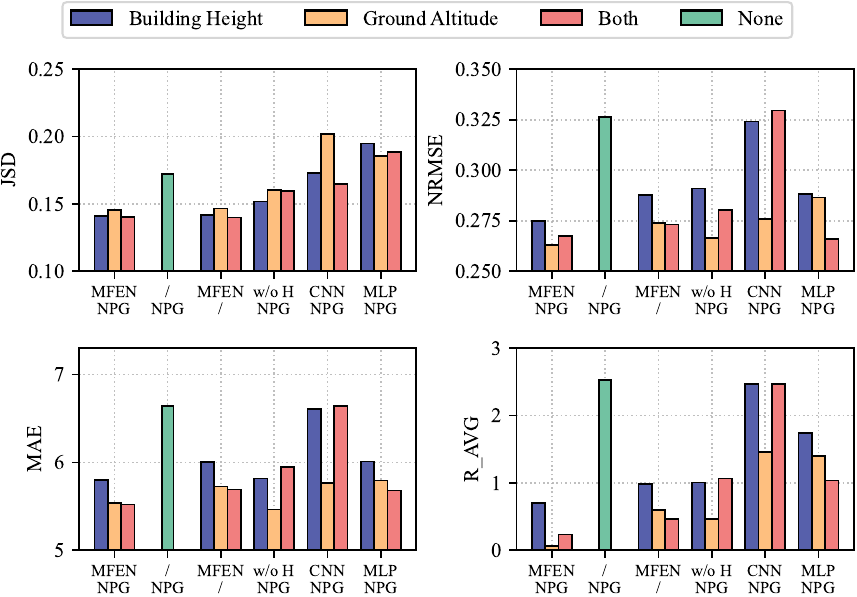}
\caption{Results of ablation studies. The performance is evaluated on the RSRP–Image dataset. The x-axis labels indicate the modules or algorithms being ablated. The first row corresponds to ablations of MFEN, and the second row corresponds to ablations of Noise Prior Guidance (NPG). The symbol “/” indicates that the module is completely removed. Specifically, w/o H indicates removing the Hessian operator from MFEN, and CNN and MLP represent replacing MFEN with simple CNN and MLP models for extracting the microenvironment embedding $\pmb{e}_\text{ME}$.}
\label{Fig_ablation}
\vspace{-4mm}
\end{figure}

In this section, we conduct ablation studies on the proposed Microenvironment Feature Extraction Network. The experiment targets four sub-questions:
\begin{itemize}
    \item i) Does incorporating urban microenvironmental features improve RSRP prediction performance?
    \item ii) Is the MFEN model effective for extracting the SS attenuation features from the multi-attribute map?
    \item iii) Does the image edge enhancement conducted by the Hessian operator provide performance gains for MFEN?
    \item iv) What is the role of the building height and ground altitude maps in MFEN?
\end{itemize}
To address these questions, we conduct ablation experiments on the MFEN architecture, with results shown in Fig.~\ref{Fig_ablation}. Here, w/o H indicates the removal of the Hessian operator, while CNN and MLP represent extracting the microenvironment embedding $\pmb{e}_\text{ME}$ from the multi-attribute map $\mathcal{M}$ using simple CNN and MLP models, respectively. We further examine changes in model performance when using only $\text{Map}_\text{BHgt}$ or $\text{Map}_\text{Alti}$ alone. This experiment is to explore the differing impacts of building distribution and ground elevation on RSRP prediction.

\begin{figure}[tb]
\centering
\includegraphics[width=0.85\linewidth]{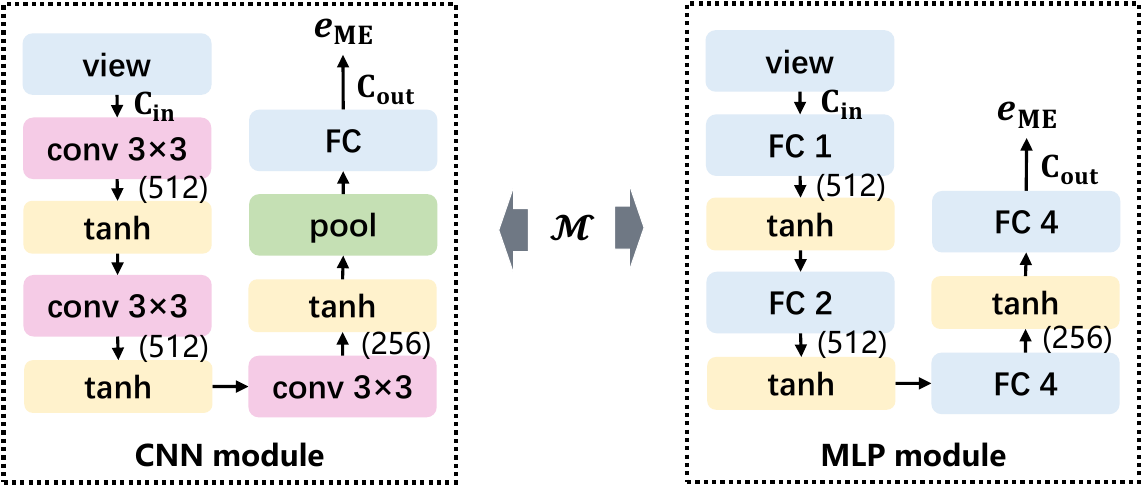}
\caption{Schematic diagrams of the CNN and MLP modules used in the ablation study.}
\label{Fig_MLP_CNN}
\vspace{-4mm}
\end{figure}

The findings from the experiments can be summarized as follows. 
{i)} According to Fig.~\ref{Fig_ablation}, the model without the MFEN module (green bars) achieves the worst performance compared with almost all other ablation settings. This indicates that urban microenvironment features have a non-negligible impact on RSRP. 
{ii)} To investigate whether the specially designed MFEN provides additional gains in effectively extracting urban microenvironment embeddings and capturing SS attenuation characteristics from the multi-attribute map, we keep the inputs and outputs unchanged and replace MFEN with simple CNN and MLP modules, as shown in Fig.~\ref{Fig_MLP_CNN}. We ensure fair comparison by aligning the intermediate feature dimensionality of MFEN, CNN, and MLP. Comparing the major columns labeled “MFEN NPG,” “CNN NPG,” and “MLP NPG” in Fig.~\ref{Fig_ablation}, the model with MFEN outperforms models with CNN or MLP modules. This indicates that MFEN is relatively more effective in capturing the SS feature influenced by the urban microenvironment. 
{iii)} Comparing the major columns labeled “MFEN NPG” and “w/o H NPG” in Fig.~\ref{Fig_ablation}, MFEN w/o H performs worse than MFEN across all four evaluation metrics. This indicates that using the Hession operator to extract and enhance edge information in $\mathcal{M}$ provides a performance gain in capturing SS-related features. The edge features in $\mathcal{M}$ are specifically related to the SS attenuation. 
{iv)} We analyse model performance using the complete multi-attribute map, using only the building-height map $\text{Map}_\text{BHgt}$, and using only the ground-elevation map $\text{Map}_\text{Alti}$. It shows that models incorporating both building-height and ground-altitude information achieve optimal or near-optimal performance in most configurations and metrics, substantially outperforming single-input models, indicating strong complementarity between the two types of information. Furthermore, models that consider only building height generally outperform those that consider only ground altitude across most configurations and metrics. This is an interesting finding, as we might intuitively expect building distributions to have a stronger impact on signal blockage and reflection. However, the experimental results indicate the opposite trend. A reasonable explanation is that LS features such as shadowing and blockage, as well as SS features, such as multipath propagation induced by building distribution, have already been sufficiently modelled and represented through Occlusion \& Shadow Modelling and Multipath Propagation Modelling. Therefore, when extracting urban microenvironment features, ground-altitude information contributes relatively greater gains.

\subsubsection{Multipath Propagation Modelling}
\begin{figure}[tb]
    \centering
    \includegraphics[width=\linewidth]{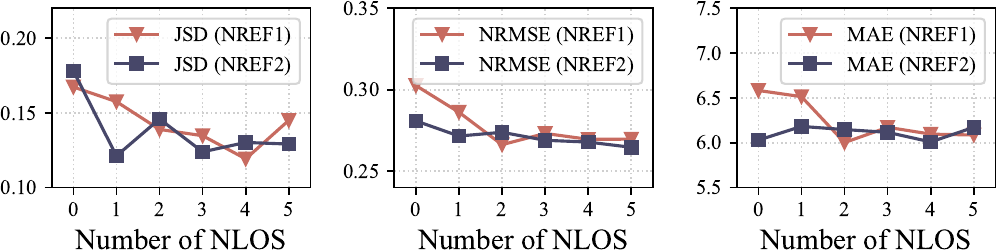}
    \caption{Performance curves of Channel-Diff under JSD, NRMSE, and MAE on RSRP–Image dataset when adjusting the number of considered NLOS and the considered reflection order. NREF1 indicates that only first-order reflections are considered, while NREF2 indicates that both first- and second-order reflections are included.}
    \label{Fig_nlos}
\end{figure}

\begin{figure}[tb]
    \centering
    \includegraphics[width=\linewidth]{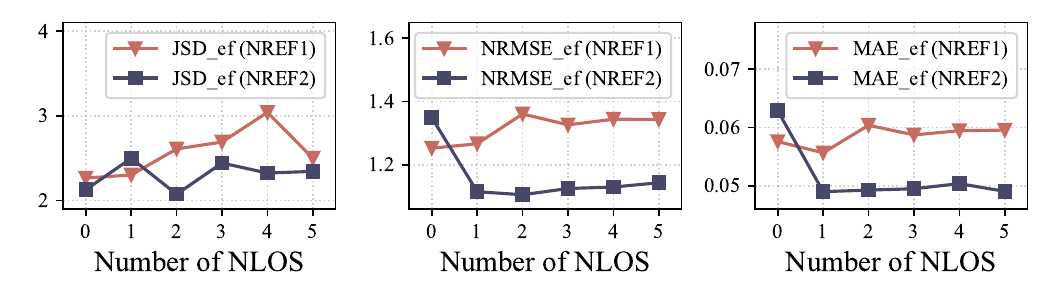}
    \caption{Performance density curves of Channel-Diff under JSD, NRMSE, and MAE on RSRP–Image dataset when adjusting the number of considered NLOS and the considered reflection order.}
    \label{Fig_nlosef}
    \vspace{-4mm}
\end{figure}

According to Section~\ref{sec::multipath}, in the physics representation layer, we construct a physical representation of signal multipath propagation, the reflection embedding $\pmb{e}_\text{RE}$, based on the polygonalized urban building map. $\pmb{e}_\text{RE}$ is a set of parameter vectors determined by Eq.~(\ref{eq:ref1}) and Eq.~(\ref{eq:ref2}), where the maximum reflection order $N_{\mathrm{ref}}$ and the number of NLOS components $N_{\mathrm{NLOS}}$ included are key parameters that affect computational cost and representation performance. Fig.~\ref{Fig_nlos} shows the impact of different numbers of rays on model performance when considering only first-order reflections (NREF1) or both first- and second-order reflections (NREF2). The results indicate that prediction accuracy is positively correlated with $N_{\mathrm{NLOS}}$. Notably, considering both first- and second-order reflections does not provide a significant improvement over considering only first-order reflections. We define the performance density $m_\text{ef} = \left[{m\log_{10}(\text{FLOPs})}\right]^{-1}$,
where $m$ denotes a lower-is-better evaluation metric, such as JSD, NRMSE, or MAE. FLOPs represent the average number of floating-point operations required to compute $\pmb{e}_\text{RE}$. A higher $m_\text{ef}$ indicates higher performance density. Fig.~\ref{Fig_nlosef} presents performance density curves measured by JSD, NRMSE, and MAE. The results show that considering more than four reflected paths contributes no further gain in performance density. Moreover, incorporating both first- and second-order reflections leads to a significant loss in performance density. This indicates that multipath effects are primarily determined by a small number of first-order reflected paths. Because reflection points absorb and scatter energy, signals from higher-order reflections have a negligible impact on the received power at the UE.

\section{Discussions}
\label{sec:Disc}

\subsection{Transferability}
This section discusses the transferability of Channel-Diff. We conduct zero-shot and few-shot experiments to verify the zero-shot and few-shot learning capabilities of the model with embedded physical knowledge when transferred to new datasets. We use the Channel-Diff model, trained through the teacher phase on the RSRP-Image dataset, and test it on the RSRP-CPGMCM dataset. The zero-shot experiment involves direct inference, while the few-shot experiment uses 5\%, 10\%, and 20\% of the RSRP-CPGMCM dataset for model training and then inference. Table~\ref{Tab_zero} shows the zero-shot learning performance and 5\% few-shot learning performance of Channel-Diff and seven other baselines. Table~\ref{Tab_few} presents the 10\% and 20\% few-shot learning performance.

From Table~\ref{Tab_zero}, it can be seen that Channel-Diff's zero-shot learning performance, or direct transfer performance, is the best overall, with a relative improvement of no less than 20.08\% compared to the baseline models. This indicates that the teacher phase training has endowed the diffusion model with universal prior knowledge applicable across various scenarios. Due to the shared knowledge, the transferring performance can be ensured. Additionally, from Table~\ref{Tab_zero} and Table~\ref{Tab_few}, it is evident that Channel-Diff demonstrates excellent few-shot learning capability. At 5\%, 10\%, and 20\% few-shot rates, the model achieves relative performance improvements of no less than 21.42\%, 17.15\%, and 21.69\%, respectively. These results confirm that the Channel-Diff model, guided by physical knowledge, possesses outstanding cross-dataset transferability and that the physical knowledge plays a crucial role in enhancing model interpretability.
\begin{table*}[tb]
\centering
\caption{Zero-shot learning performance and 5\% few-shot learning performance of Channel-Diff and other 7 baselines.}
\setlength{\tabcolsep}{3pt}
\begin{tabular}{cccccccccccccccc}
\toprule
\multirow{3}{*}{Method} & \multicolumn{6}{c}{RSRP-Image $\rightarrow$ RSRP-CPGMCM (zero-shot)} & \multirow{3}{*}{$\bar{\Delta}$} & \multicolumn{6}{c}{RSRP-Image $\rightarrow$ RSRP-CPGMCM (5\% few-shot)} & \multirow{3}{*}{$\bar{\Delta}$} \\
\cmidrule(lr){2-7} \cmidrule(lr){9-14}
 & JSD$\downarrow$ & $\Delta$ & NRMSE$\downarrow$ & $\Delta$ & MAE$\downarrow$ & $\Delta$ & & JSD$\downarrow$ & $\Delta$ & NRMSE$\downarrow$ & $\Delta$ & MAE$\downarrow$ & $\Delta$ & \\
\midrule
MLR & 0.8319 & 86.85\% & 1.076 & 70.25\% & 36.82 & 73.50\% & \textbf{76.87\%} & 0.3659 & 73.64\% & 0.2981 & 1.88\% & 8.881 & 2.35\% & \textbf{25.96\%} \\
\midrule
WPM-BPNN & 0.3771 & 70.99\% & 0.3527 & 9.24\% & 10.71 & 8.88\% & \textbf{29.70\%} & 0.3791 & 74.56\% & 0.2587 & -13.07\% & 8.918 & 2.76\% & \textbf{21.42\%} \\
LSTM-LQE & 0.7572 & 85.55\% & 0.349 & 8.28\% & 9.765 & 0.06\% & \textbf{31.30\%} & 0.7551 & 87.23\% & 0.3707 & 21.10\% & 10.01 & 13.37\% & \textbf{40.56\%} \\
VAE+DNN & 0.4256 & 74.30\% & 0.3387 & 5.49\% & 10.22 & 4.51\% & \textbf{28.10\%} & 0.4275 & 77.44\% & 0.2883 & -1.46\% & 8.745 & 0.83\% & \textbf{25.60\%} \\
\midrule
MTS-CGAN & 0.4111 & 73.39\% & 0.3243 & 1.30\% & 9.307 & -4.86\% & \textbf{23.28\%} & 0.4663 & 79.31\% & 0.2998 & 2.43\% & 8.917 & 2.75\% & \textbf{28.17\%} \\
CSDI & 0.7016 & 84.41\% & 0.6983 & 54.16\% & 22.55 & 56.72\% & \textbf{65.10\%} & 0.5364 & 82.02\% & 0.5273 & 44.53\% & 16.13 & 46.24\% & \textbf{57.59\%} \\
CANDLE & 0.2957 & 63.00\% & 0.3299 & -0.06\% & 9.503 & -2.69\% & \textbf{20.08\%} & 0.2761 & 65.06\% & 0.2941 & 0.54\% & 8.715 & 0.49\% & \textbf{22.03\%} \\
\midrule
\textbf{Channel-Diff} & 0.1094 & - & 0.3201 & - & 9.759 & - & - & 0.0965 & - & 0.2925 & - & 8.672 & - & - \\
\bottomrule
\end{tabular}
\begin{center}
    Hint: $\downarrow$ means lower is better. $\bar\Delta$ represents the average percentage improvement. Bold numbers are all the positive average percentage improvements.
\end{center}
\label{Tab_zero}
\end{table*}

\begin{table*}[tb]
\centering
\caption{10\% few-shot learning performance and 20\% few-shot learning performance of Channel-Diff and the other 7 baselines.}
\setlength{\tabcolsep}{3pt}
\begin{tabular}{cccccccccccccccc}
\toprule
\multirow{3}{*}{Method} & \multicolumn{6}{c}{RSRP-Image $\rightarrow$ RSRP-CPGMCM (10\% few-shot)} & \multirow{3}{*}{$\bar{\Delta}$} & \multicolumn{6}{c}{RSRP-Image $\rightarrow$ RSRP-CPGMCM (20\% few-shot)} & \multirow{3}{*}{$\bar{\Delta}$} \\
\cmidrule(lr){2-7} \cmidrule(lr){9-14}
 & JSD$\downarrow$ & $\Delta$ & NRMSE$\downarrow$ & $\Delta$ & MAE$\downarrow$ & $\Delta$ & & JSD$\downarrow$ & $\Delta$ & NRMSE$\downarrow$ & $\Delta$ & MAE$\downarrow$ & $\Delta$ & \\
\midrule
MLR & 0.3974 & 73.20\% & 0.2932 & 1.60\% & 8.548 & 0.11\% & \textbf{24.97\%} & 0.4485 & 76.34\% & 0.2779 & 0.76\% & 8.366 & 3.19\% & \textbf{26.76\%} \\
\midrule
WPM-BPNN & 0.3681 & 71.07\% & 0.2584 & -11.65\% & 7.909 & -7.97\% & \textbf{17.15\%} & 0.3988 & 73.40\% & 0.2591 & -6.45\% & 7.95 & -1.87\% & \textbf{21.69\%} \\
LSTM-LQE & 0.7455 & 85.71\% & 0.3764 & 23.35\% & 10.18 & 16.12\% & \textbf{41.73\%} & 0.743 & 85.72\% & 0.373 & 26.06\% & 10.11 & 19.89\% & \textbf{43.89\%} \\
VAE+DNN & 0.3189 & 66.60\% & 0.2985 & 3.35\% & 8.901 & 4.07\% & \textbf{24.67\%} & 0.2614 & 59.41\% & 0.2954 & 6.64\% & 8.817 & 8.14\% & \textbf{24.73\%} \\
\midrule
MTS-CGAN & 0.4666 & 77.18\% & 0.362 & 20.30\% & 10.52 & 18.83\% & \textbf{38.77\%} & 0.4365 & 75.69\% & 0.3307 & 16.60\% & 9.666 & 16.21\% & \textbf{36.17\%} \\
CSDI & 0.4881 & 78.18\% & 0.4804 & 39.95\% & 14.52 & 41.19\% & \textbf{53.11\%} & 0.5296 & 79.97\% & 0.5185 & 46.81\% & 15.84 & 48.87\% & \textbf{58.55\%} \\
CANDLE & 0.2677 & 60.22\% & 0.295 & 2.20\% & 8.707 & 1.93\% & \textbf{21.45\%} & 0.248 & 57.22\% & 0.2961 & 6.86\% & 8.734 & 7.27\% & \textbf{23.78\%} \\
\midrule
\textbf{Channel-Diff} & 0.1065 & - & 0.2885 & - & 8.539 & - & - & 0.1061 & - & 0.2758 & - & 8.099 & - & - \\
\bottomrule
\end{tabular}
\begin{center}
    Hint: $\downarrow$ means lower is better. $\bar\Delta$ represents the average percentage improvement. Bold numbers are all the positive average percentage improvements.
\end{center}
\label{Tab_few}
\end{table*}

\subsection{The Field of View (FoV) in MFEN}
\begin{figure}[tb]
    \centering
    \includegraphics[width=\linewidth]{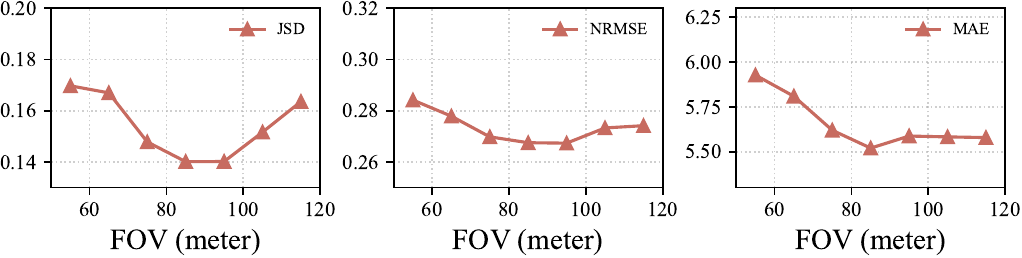}
    \caption{Performance curves of Channel-Diff under JSD, NRMSE, and MAE on RSRP–Image dataset when adjusting the FoV $l$.}
    \label{Fig_FOV}
    \vspace{-4mm}
\end{figure}

In this section, we discuss the impact of a key parameter in MFEN, the FoV $l$, on model performance. FoV represents the visible edge length that the Serializer $\mathcal{S}_\text{cut}$ in MFEN crops the urban microenvironment around the UE. Fig.~\ref{Fig_FOV} shows how model performance changes as $l$ varies. It is observed that when $l$ increases within the range [50, 120] meters, the model performance first improves and then decreases. The optimal performance occurs at $l \approx 85$ meters. This indicates that SS fading is influenced by the urban microenvironment within a certain spatial range. If the microenvironment map covers too small an area, the model cannot sufficiently capture the relevant context around the UE; if it covers too large an area, it introduces urban information that does not affect the UE, making it harder for the model to focus on the effective microenvironmental region.

\subsection{Sensitivity of the Series Length}
\begin{figure}[tb]
    \centering
    \includegraphics[width=\linewidth]{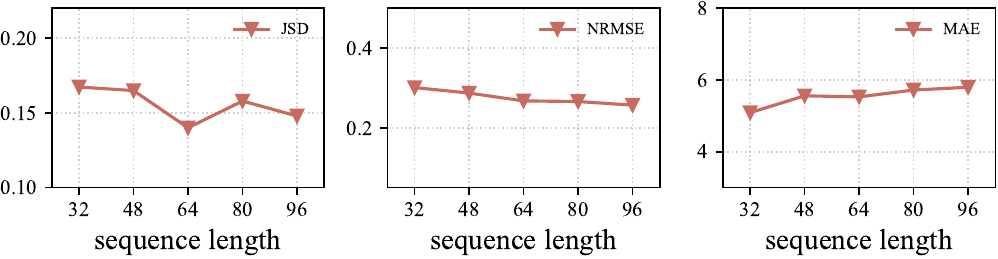}
    \caption{Performance curves of Channel-Diff under different settings of the length of the RSRP series.}
    \label{Fig_length}
    \vspace{-4mm}
\end{figure}

This paper formulates RSRP prediction as a time-series generation task. Channel-Diff supports the generation of RSRP sequences of arbitrary length, provided that aligned external condition information is available. In this section, we investigate whether different sequence length settings affect average prediction accuracy. Fig.~\ref{Fig_length} varies the RSRP sequence length from 32 to 96. The results show that model performance is highly stable and that JSD and NRMSE performance exhibit slight improvements as sequence length increases. This may be attributed to intrinsic temporal dependencies within longer sequences, which enhance the model’s estimation capability. Ignoring this phenomenon, the experiment demonstrates that Channel-Diff is insensitive to sequence length and can be applied to RSRP prediction tasks with varying sequence lengths.

\subsection{Convergence Efficiency Analysis}
In this section, we discuss the effectiveness of the two-stage training scheme in terms of training convergence efficiency. As mentioned before, the pretraining in the teacher stage initializes the model parameters using theoretically calculated RSRP values, guiding the model toward an LS physically interpretable direction. The retraining process in the student stage builds upon the teacher stage, further extracting the complex environmental features and multi-scale attenuation characteristics exhibited in real data. This process achieves the infusion of physical knowledge into the diffusion model. Since the physical knowledge embedded in the propagation model primarily pertains to the LS attenuation characteristics of signals, a more thorough infusion of physical knowledge enhances the model's ability to estimate LS attenuation. Consequently, when trained on real data that includes both LS and SS scales, the model can "focus more" on extracting SS attenuation characteristics. This aligns with the principle of "sharpening the axe before chopping wood", which means moderately increasing the training duration in the teacher stage improves training efficiency in the student stage.
\begin{figure}[tb]
\centering
\includegraphics[width=0.9\linewidth]{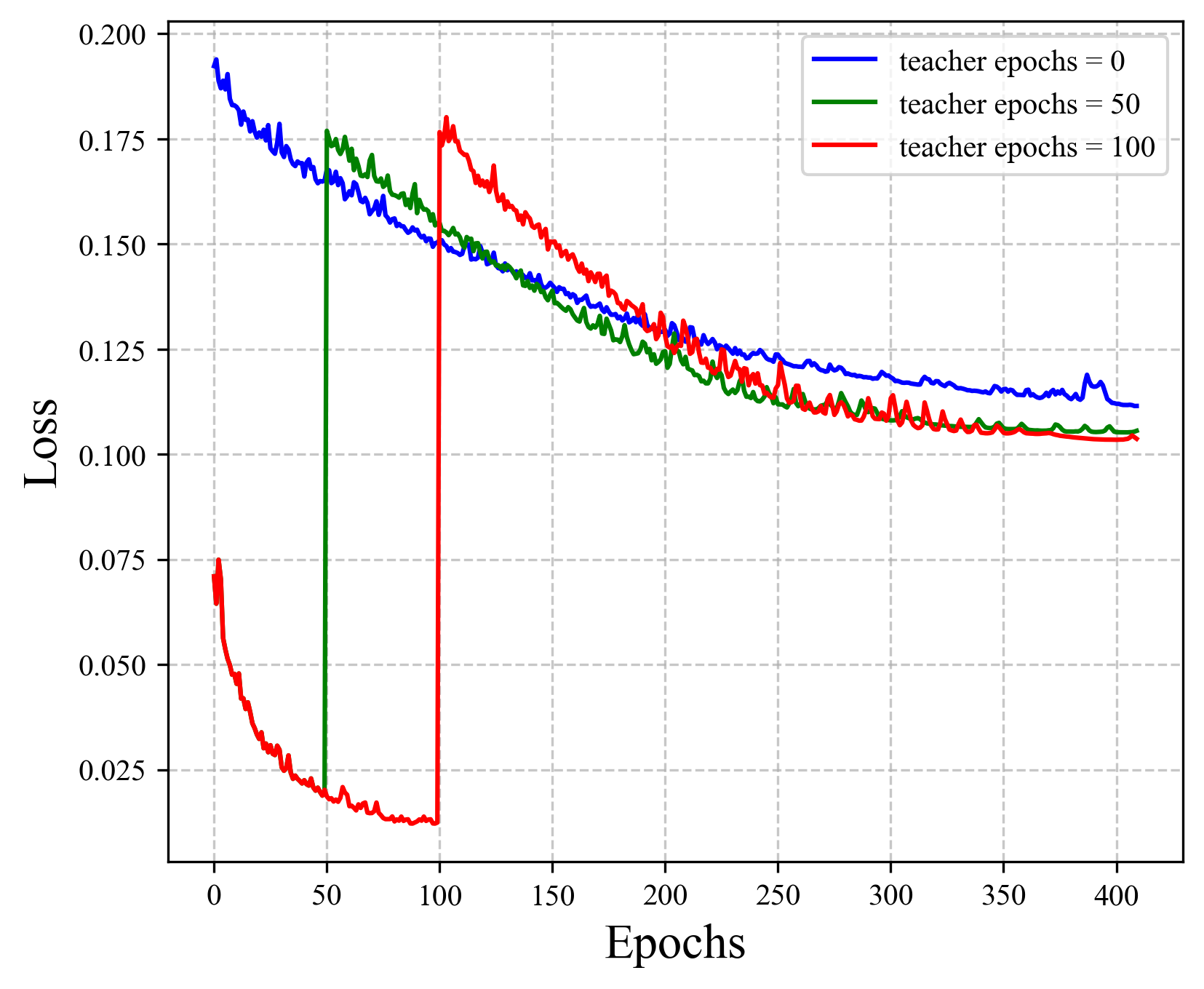}
\caption{Training loss curves with different numbers of teacher epochs on the RSRP-CPGMCM dataset.}
\label{Fig_loss}
\vspace{-4mm}
\end{figure}

Fig.~\ref{Fig_loss} shows the training error curves on the RSRP-CPGMCM dataset when the number of training iterations in the teacher stage is set to 0, 50, and 100, respectively. There is a noticeable loss-jump between the teacher and student phases. This is because the training set switches from $\textbf{RSRP}_\text{calc}$ to $\textbf{RSRP}_\text{real}$. While $\textbf{RSRP}_\text{calc}$ is easier to fit, $\textbf{RSRP}_\text{real}$ contains more complex influencing factors. It is evident that as the teacher stage training increases, the initial loss value at the beginning of the student training stage decreases significantly. This indicates that moderate teacher-stage training enhances the model's ability to estimate RSRP on real data. Additionally, a more important observation is that with increased teacher-stage training, the model exhibits a steeper loss reduction trend in the student stage. The red curve, which starts the student stage training the latest, reaches its lowest loss level at around 400 iterations. This phenomenon strongly demonstrates that the incorporation of physical knowledge facilitates the model's feature extraction on real data. The proposed teacher-student training paradigm enhances training efficiency in the RSRP estimation task.

\subsection{Future Works}
Channel-Diff provides operators with an effective tool for channel quality estimation and link prediction. However, an important issue that cannot be ignored when deploying this model in live networks is the accuracy of UE location information. Since reporting UE GPS positions introduces additional overhead, operators typically record the location of the serving BS as a proxy for the UE location, which can result in errors of tens or even hundreds of meters. This can significantly affect downstream applications and optimisation algorithms that rely on user location. This work is based on accurately recorded UE locations. In future work, we will investigate how to perform channel estimation and link prediction using coarsely reported user locations, which will improve the generalisability of the approach and broaden its application scenarios.

\section{Conclusion}
\label{conclusion}
This paper proposes Channel-Diff, a diffusion-based physics-informed generative model for RSRP prediction. By integrating multi-scale channel characteristics and prior physical knowledge into a conditional diffusion framework, Channel-Diff successfully bridges the gap between traditional propagation models and modern data-driven approaches. The model captures both LS and SS attenuation effects through physics-aligned conditioning, multi-modal feature embeddings, and a physics-prior-guided two-stage training paradigm. Experiments conducted on two real-world datasets demonstrate that Channel-Diff achieves at least 25.15\% to 37.19\% performance improvements over state-of-the-art baselines, while offering enhanced interpretability, stronger generalization across scenarios, and improved training efficiency. These results highlight the potential of physics-informed generative models to serve as reliable tools for wireless channel prediction and network optimization, paving the way for more adaptive, accurate, and efficient communication systems.


\vspace{-2mm}
\bibliographystyle{IEEEtran}
\bibliography{Ref}

\vfill
\end{document}